\documentclass{article}

\usepackage{arxiv}

\usepackage{xspace}
\usepackage[utf8]{inputenc} 
\usepackage[T1]{fontenc}    
\usepackage{hyperref}       
\usepackage{url}            
\usepackage{booktabs}       
\usepackage{amsfonts}       
\usepackage{nicefrac}       
\usepackage{microtype}      
\usepackage{lipsum}		
\usepackage{graphicx}
\usepackage{doi}
\usepackage{algorithm}
\usepackage{algpseudocode}
\usepackage{multirow}
\usepackage{multicol}
\usepackage{subfig}
\usepackage{amsmath}
\usepackage{upgreek}
\usepackage{mathtools}
\DeclarePairedDelimiter\abs{\lvert}{\rvert}

\scshape

\newcommand{\hww}{\ensuremath{\mathrm{H}\to\mathrm{W^+}\mathrm{W^-}}\xspace}

\title{Model independent measurements of Standard Model cross sections with Domain Adaptation}

\author{Benedetta Camaiani$^{1,2}$$^*$\\\and Roberto Seidita$^{1,2}$\\\and Lucio Anderlini$^2$\\\and Rudy Ceccarelli$^2$\\\and Vitaliano Ciulli$^{1,2}$\\\and Piergiulio Lenzi$^{1,2}$\\\and Mattia Lizzo$^{1,2}$\\\and Lorenzo Viliani$^2$}

\date{%
    $^1$Department of Physics and Astronomy, Università degli Studi di firenze, Via G. Sansone 1, Sesto Fiorentino, 50019, Firenze, Italy.\\%
    $^2$Sezione di Firenze, Istituto Nazionale di Fisica Nucleare, Via G. Sansone 1, Sesto Fiorentino, 50019, Firenze, Italy.\\%
    $^*$Corresponding author\\[2ex]%
    \today
}

\renewcommand{\headeright}{}
\renewcommand{\undertitle}{}

\begin{document}

\maketitle

\begin{abstract}
    With the ever growing amount of data collected by the ATLAS and CMS experiments at the CERN LHC, fiducial and differential measurements of the Higgs boson production cross section have become important tools to test the standard model predictions with an unprecedented level of precision, as well as seeking deviations that can manifest the presence of physics beyond the standard model. These measurements are in general designed for being easily comparable to any present or future theoretical prediction, and to achieve this goal it is important to keep the model dependence to a minimum. Nevertheless, the reduction of the model dependence usually comes at the expense of the measurement precision, preventing to exploit the full potential of the signal extraction procedure. In this paper a novel methodology based on the machine learning concept of domain adaptation is proposed, which allows using a complex deep neural network in the signal extraction procedure while ensuring a minimal dependence of the measurements on the theoretical modelling of the signal.
\end{abstract}


\maketitle

\section{Introduction}
High energy physics (HEP) experiments report their results in a number of different ways, such as showing exclusion limits of particular theoretical models, presenting the measured production cross section of the process of interest, or measuring parameters of a given model. Different ways of reporting the results are usually characterized by different degrees of underlying assumptions, that can make the measurements more (less assumptions) or less (more assumptions) usable for re-interpretations.

Measurements of fiducial differential production cross sections are usually designed to minimize the underlying model assumptions, such that they can be easily re-interpreted to set constraints on, in principle, any theoretical model. Typically these are measurements of cross section in bins of the observable of interest (differential) with limited extrapolation to a restricted portion of the phase space (fiducial volume) at particle-level, i.e. prior to the simulation of the interaction with the detector. This phase space is defined to match as closely as possible the experimental event selection, so that the extrapolation introduces as little model dependence as possible (certainly much less than what would arise from the assumptions one would have to introduce if the measurement was reported in the full phase space).

These measurements are of particular interest in the fields of standard model (SM) and Higgs boson physics at the CERN Large Hadron Collider (LHC), and the ATLAS and CMS experiments already published many measurements of this kind.

In the case of Higgs boson cross section measurements, a complementary approach was formulated in the past few years, known as the Simplified Template Cross Section (STXS) framework~\cite{Berger:2019wnu}. This approach consists in the measurement of production cross sections in pre-defined bins of particle-level phase space, established with the goal of minimizing the measurement dependence on theoretical assumptions, isolating possible BSM effects, and easing the comparison and combination of results from different experiments. No fiducial region concerning the Higgs decay is involved in the STXS phase space definition, except for a loose requirement on the Higgs boson rapidity ($\abs{ y_\mathrm{H}}<2.5$). This allows to combine measurements in different channels at the expense of somewhat larger extrapolation uncertainties with respect to those one would be able to achieve with a fiducial phase space specifically tailored to the event selection.

For both the fiducial differential and STXS approaches, it is crucial to perform an accurate evaluation of the sources of bias in the measurement arising from the underlying theoretical model assumptions. Typically, the analysis strategy is devised using Monte Carlo (MC) simulations of the SM signal processes as benchmark models, which could lead to a biased result if one wanted to re-interpret the measurement in terms of beyond-the-SM (BSM) signal hypotheses.

In some cases the model assumptions can introduce a significant amount of bias, therefore spoiling the re-interpretability of the measurements. 
This is especially relevant when complex machine learning (ML) techniques are used in the analyses. In fact, it is common nowadays to develop powerful ML algorithms to improve the measurement precision, for example by maximizing the signal-to-background separation in classification problems, exploiting complex deep neural network (DNN) architectures. The improvements one obtains using these tools in the signal extraction procedure however often come at the expense of larger model assumptions, as explained in Section~\ref{sec:mod_dep}.

In this paper a novel methodology based on the concept of domain adaptation (DA) of a ML algorithm is proposed. This allows using complex DNNs in the signal extraction procedure while ensuring a minimal dependence of the measurements on the theoretical modeling of the signal. This can be achieved by expanding the DNN architecture with a second adversary DNN that makes the full algorithm model-agnostic, as described in detail in Section~\ref{sec:ADNN}.

In Section~\ref{sec:hww} this approach is applied to the use case of the \hww STXS cross section measurements at the LHC, showing that the proposed method can drastically reduce the model dependence uncertainties in the measurements, without significantly compromising the precision of the results.

\section{Model dependence of fiducial and differential cross section measurements}\label{sec:mod_dep}

Differential and fiducial cross section measurements represent a valuable tool for probing alternative hypotheses with respect to the SM. For this reason it is desirable to limit the model dependence of the analysis, providing a result with as little bias as possible towards the SM expectation. 
Nevertheless, some sources of model dependence often arise from the analysis strategy itself. 

The main one lies in the signal extraction procedure, especially when it is performed via a template fit. The basic idea is to extract the probability density function (PDF) of a discriminating observable, for both signal and background, from the MC simulation and then fit the sum of those PDFs, along with the signal and background normalizations, to data. The problem arises from the fact that the shape of the observable distribution to be fitted to data may in general depend on the properties of the physics model governing the signal process under consideration. For instance, the Higgs quantum numbers, like CP-parity,  can be predicted to be different from the SM ones by several BSM theories, with non negligible consequences on the shapes of several observables.
Ideally, one would employ a model agnostic variable with high capability to discriminate the signal of interest against the background as a fit variable.
In some measurements, however, a model independent variable can not be defined in a straightforward fashion. For instance, this may be due to the failure in detecting all particles produced in proton-proton (p-p) collisions, either because of inefficiencies of the experimental setup or as an inherent consequence of the considered final state (e.g, neutrinos escaping detection).

Even the usage of the output of an NN as a fit variable, which is common in HEP analyses, leads to the same consideration. Although a properly trained NN discriminant usually outperforms simpler observables in signal separation, it is still trained on sets of events obtained from MC simulation, and therefore the shape of the discriminator depends on the underlying physics hypothesis used to generate the training set, especially for the signal.\\
Therefore, these features lead to a bias in the signal extraction procedure, which in turn induces in the measurement an undesirable degree of dependence on the signal hypothesis used to generate the template.

The unfolding procedure~\cite{Kuusela_2015}, necessary for taking into account the smearing introduced by the detector and for the the extrapolation of the result to the fiducial region, can also be a source of possible bias. The response matrix quantifying the smearing between the reconstructed- and particle-level phase space may indeed depend on the physics model. In case the response matrix is quite diagonal, however, the dependence is expected to be small and can be neglected.

Finally, another contribution originates from the definition of the analysis phase space itself. In fact, the event selection efficiency, as well as the acceptance factor, may be different depending on the considered signal model. The former refers to the probability of reconstructing a signal event in the fiducial volume, and hence depends on the detector smearing, whereas the latter can be estimated a posteriori using only particle-level quantities and therefore be accounted for.

The approach presented in this study is focused on the reduction of the model dependence originating from the signal extraction procedure as described in Section~\ref{sec:ADNN}. The other sources of model dependence described above are usually tackled by means of different approaches and are not covered by this study.

\section{Model dependence reduction as a problem of Domain Adaptation}\label{sec:ADNN}
In order to preserve the model independence of the differential and fiducial analyses, it is necessary to define a learning algorithm that classifies events correctly and at the same time does not significantly distinguish different physics models assumed to generate signal events.\\ 
The latter goal generally falls within the realm of what is usually defined as domain adaptation (DA)~\cite{article_DA}, a branch of ML concerned with studying how an algorithm performs when evaluated on a data set statistically different from the one it was trained on. Several techniques have been proposed to allow the construction of DNNs whose performance is robust when applied on a range of different data sets, called domains~\cite{10.5555/2946645.2946704}~\cite{https://doi.org/10.48550/arxiv.1611.01046}. 

In the use case of \hww STXS measurements the different data sets correspond to events generated assuming different signal models and the DA is achieved if the NN is agnostic with respect to the considered signal hypothesis, which in practice means that the shape of the output discriminant is the same regardless of the model assumption.

\subsection{Implementation of an adversarial deep neural network}\label{sec:ADNN_implementation}
\begin{figure*}[t]
    \centering
    {\includegraphics[width=0.8\columnwidth]{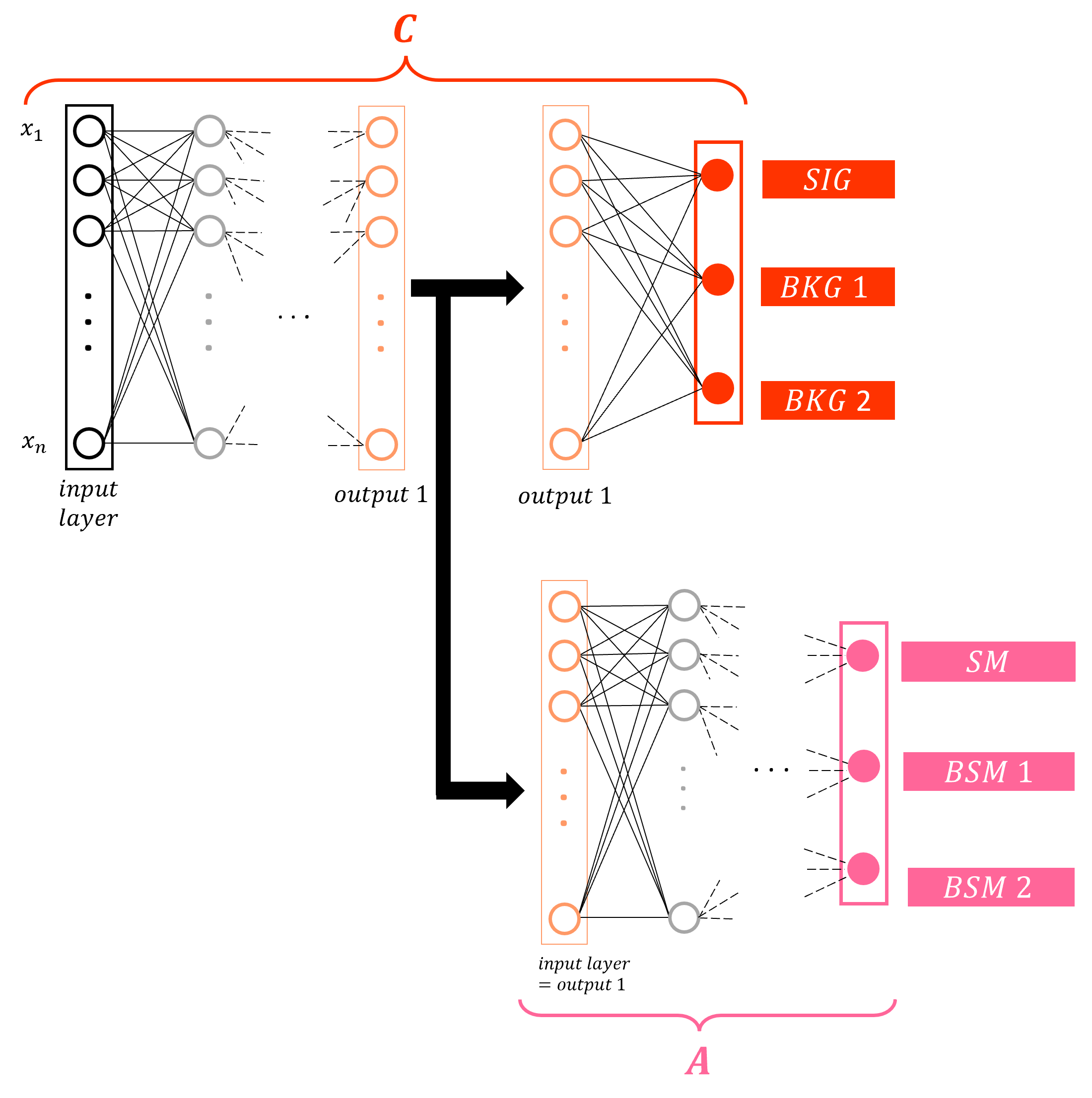}} 
    \caption{\emph{Schematic view of the adversarial deep neural network.}}
    \label{ADNN}
    \end{figure*}

The implementation of DA developed in this work is based on an adversarial deep neural network (ADNN). The ADNN is a system of two networks, consisting of a classifier ($C$) and an adversary ($A$), which are trained in a competitive way to perform different tasks:
the classifier aims to determine if an event is signal- or background-like, and is trained on a data sample including events arising from different domains i.e., different signal models. On the other hand, the goal of the adversary is to guess the physics model of a signal event, regressing the domain from the second-to-last layer of the classifier.\\ The two networks are trained so that $C$ learns to discriminate between signal and background, but is penalized if the optimized data representation contains too much information on the domain of origin of signal events. This training approach fosters the emergence of features among the classifier input variables that provide discriminating power for the main learning task (signal-to-background separation) while not relying on the variation in inputs due to the different signal models.
This goal has been achieved by implementing a two-step training procedure on a labeled data sample described in the following.\\ 
In each epoch the classifier is first trained with a combined loss function, defined as \begin{equation}
    \mathcal{L}(C+A) = \mathcal{L}(C)-\alpha\cdot\mathcal{L}(A)
\end{equation}
where $\mathcal{L}(C)$ is the categorical cross-entropy (CCE) loss function of the classifier in the event it has to discriminate the signal from more than one background process, $\mathcal{L}(A)$ is the CCE loss function of the adversary and $\alpha$ is a configurable hyperparameter that regulates how much the network is penalized for learning the signal model. During this preliminary step the weights of $A$  are kept frozen and the minimization is performed only with respect to the weights of $C$.\\
Subsequently, in the same epoch, the adversary is trained with the $\mathcal{L}(A)$ loss function. Algorithm~\ref{alg_training} schematizes the training procedure.\\
The hyperparameters defining the ADNN need to be optimized so that $C$ is able to discriminate events while simultaneously preventing $A$ from inferring the domain. When, and if, an equilibrium between the performance of each of the networks is reached, the output of the classifier is independent of the signal model.\\ The general structure of an ADNN with three output nodes for both the classifier and the adversary is represented as an example in Figure~\ref{ADNN}: the classifier takes as input a set of $m$ features $\{x_i\}$ and has to discriminate the signal from two possible background processes; finally, its second-to-last layer (commonly referred to as \emph{representation}) constitutes the input layer of the adversary, which is trained to predict the model used to generate each training event, randomly chosen among the SM signal hypothesis and a number of possible BSM scenarios. In principle, the alternative scenarios should cover the full spectrum of BSM models that are relevant.\\
The structures of the classifier and the adversary can be defined using different hyperparameter values: the architecture of the entire ADNN needs to be optimized for both separation of signal from background and decorrelation of the classifier output scores from the domain to which signal events belong, as will be described in Section~\ref{sec:opt}.
Such optimization can be accomplished by maximizing the categorical accuracy of the classifier and by simultaneously minimizing the two-sample Kolmogorov-Smirnov (K-S) test statistic between the distributions of the classifier when evaluated on signal events simulated under the different considered hypotheses. The categorical accuracy is the fraction of total events correctly classified and, for a model trained on a dataset composed of the same number of events for each class, it is directly related to diagonal elements of the confusion matrix. When an imbalanced datasets is used, this assumption does not hold and other metrics based on confusion matrix can be employed for evaluating the performance.  The K-S test statistic provides a measure of compatibility, quantifying the distance between two distributions. Therefore one may minimize the objective function defined as the average of the K-S test statistic computed between the classifier output shapes of signal events simulated under the SM and each of the considered alternative hypotheses. The objective function defined in this way will be referred to as the average K-S test statistic in the following. If no residual model dependence is left in the classifier outputs, the distributions of signal events generated under different model assumptions are expected to be compatible within the statistical accuracy.
By doing so, the best estimation of the hyperparameters outlining the ADNN can be found.
\begin{algorithm}
\caption{Two-step training of the ADNN.}\label{alg_training}
\begin{algorithmic}
\Require Parameter vector of the classifier at epoch $k$ $\boldsymbol{\theta}\mathrm{_C^{(k)}}$
\Require Parameter vector of the adversary at epoch $k$ $\boldsymbol{\theta}\mathrm{_A^{(k)}}$
\Require Learning rate of the classifier $\eta^\mathrm{C}$
\Require Learning rate of the adversary $\eta^\mathrm{A}$
\\
\For {$\mathrm{k}$ epochs}\\
   \ \ \ with $\boldsymbol{\theta}\mathrm{_A^{(k)}}$ fixed,
   \State $g\mathrm{_C^{(k)}} \gets \nabla\mathrm{_{\theta_C}} \mathcal{L}(C+A)$;\\
    \ \ \ with $\boldsymbol{\theta}\mathrm{_C^{(k)}}$ fixed,
    \State $g\mathrm{_A^{(k)}} \gets \nabla\mathrm{_{\theta_A}} \mathcal{L}(A)$;\\

    \State $\boldsymbol{\theta}\mathrm{_C^{(k+1)}} \gets \boldsymbol{\theta}\mathrm{_C^{(k)}} - \eta\mathrm{^C} g\mathrm{_C^{(k)}}$;\\ 
    \State $\boldsymbol{\theta}\mathrm{_A^{(k+1)}} \gets \boldsymbol{\theta}\mathrm{_A^{(k)}} - \eta\mathrm{^A} g\mathrm{_A^{(k)}}$;\\ 
\EndFor
\end{algorithmic}
\end{algorithm}

\section{The \texorpdfstring{H$\rightarrow$W$^+$W$^-$}{H->WW} case}\label{sec:hww}

In this section the application of the adversarial method to the case of the \hww STXS cross section measurement at the LHC is shown, as a suitable case study where all the issues connected with the model dependence are present. It can be seen as a simplification of a possible SM analysis, performed purely at particle-level on a simulated data sample from p-p collisions at a centre of mass energy of 13 TeV, corresponding to an integrated luminosity of 138 fb$^{-1}$.\\
The analysis targets events in which a Higgs boson is produced via vector boson fusion (VBF) and subsequently decays to a pair of opposite-sign W bosons, each decaying in turn to an electron or muon and a neutrino. \\
The presence of neutrinos in the final state makes the full kinematic reconstruction, and thus the measurement of the invariant mass of the system, impossible, thus removing the possibility of using a readily available model independent discriminating variable.\\
The VBF mechanism is interesting because a precision measurement of its cross section allows to put tight constraints on the Higgs boson couplings to vector bosons and thus is an important test of the SM predictions. The HWW and, more generally, the HVV vertex is indeed predicted to have a different CP quantum number by several BSM theories, which introduce anomalous couplings between vector bosons and the Higgs particle. These models are summarized by the following equation, which represents the general scattering amplitude describing the interaction between a spin-0 Higgs boson (H) and two spin-1 gauge bosons ($\mathrm{V_1 V_2}$):
\begin{equation} 
    \mathcal{A} (\mathrm{HVV})\  \sim \ 
        a\mathrm{^{VV}_1} m^2_{\mathrm{V_1}}\epsilon^*_{\mathrm{V_1}}\epsilon^*_{\mathrm{V_2}}\ + \   \left[\frac{k\mathrm{^{VV}_1} q\mathrm{^2_{V_1}}+k\mathrm{^{VV}_2} q\mathrm{^2_{V_2}}}{(\Lambda^{\mathrm{VV}}_1)^2}\right]
    m^2_{\mathrm{V_1}}\epsilon^*_{\mathrm{V_1}}\epsilon^*_{\mathrm{V_2}}\ +\ a^{\mathrm{VV}}_2 f^{*(1)}_{\upmu\upnu}f^{*(2),\upmu\upnu}\ +\ a^{\mathrm{VV}}_3 f^{*(1)}_{\upmu\upnu}\tilde{f}^{*(2),\upmu\upnu}
\end{equation}
where $f^{(\mathrm{i})\upmu\upnu} = \epsilon^\upmu_{\mathrm{V_i}}q^\upnu_{\mathrm{V_i}} - \epsilon^\upnu_{\mathrm{V_i}}q^\upmu_{\mathrm{V_i}}$  and $f^{(\mathrm{i})}_{\upmu\upnu} = \frac{1}{2} \epsilon_{\upmu \upnu \uprho \upsigma} f^{(\mathrm{i}), \uprho \upsigma}$ are the field strength tensor and the dual field strength tensor of a gauge boson with momentum $q_{\mathrm{V_i}}$, polarization $\epsilon_{\mathrm{V_i}}$ and pole mass $m_{\mathrm{V_i}}$ \cite{Gritsan_2020}.
$\Lambda\mathrm{^{VV}_1}$ is the energy scale of the BSM physics and is a free parameter of the model. For the sake of clarity, let $L_1$ be \begin{equation}
    L_1\ \stackrel{def}{=}\  \frac{k\mathrm{^{VV}_1} q\mathrm{^2_{V_1}}+k\mathrm{^{VV}_2} q\mathrm{^2_{V_2}}}{(\Lambda^{\mathrm{VV}}_1)^2}.   
\end{equation} 
The leading order (LO) SM-like contribution corresponds to $a^{\mathrm{VV}}_1 = 1$ with VV = ZZ, WW, and $L_1, a_\mathrm{2}, a_3 = 0$. These are CP-even interactions. There can not be LO couplings to massless gauge bosons, so there are no contributions from photons and gluons at this order.
Any other VV couplings are ascribed to anomalous couplings, which can be either small SM corrections due to loop effects or new BSM phenomena. In particular, $a^{\mathrm{VV}}_2 = 1$ describes loop-induced CP-even couplings (like HZ$\mathrm{\gamma}$, H$\mathrm{\gamma\gamma}$ and Hgg), which are parametrically suppressed by the coupling constants $\alpha$ and $\alpha_\mathrm{s}$. These phenomena are described by the model which will be referred to as the H0PH physics model in the following.\\
The $L_1$ term is associated to CP-even HVf$\bar{\text{f}}$ and Hf$\bar{\text{f}}$f$\bar{\text{f}}$ interactions, which are predicted by the physics model denominated H0L1.
\\ Finally, $a^{\mathrm{VV}}_3 = 1$ accounts for three loop-induced CP-odd coupling in the SM.
This kind of processes is identified by the so-called H0M model.\\
The $a^{\mathrm{VV}}_1$ coupling is assumed to be zero by the H0PH, H0L1 and H0M models.\\
In addition, three physics theories are defined as mixtures between the SM and one of the previous BSM hypothesis. Such models correspond to having $a^{\mathrm{VV}}_1$ and one of the BSM operators both equal to $0.5$ and will be labeled by adding the \lq\lq f05\rq\rq\ tag to the model names introduced above.

The measurement is performed within the STXS phase space regions, defined at particle-level. 
For what the VBF production mechanism is concerned, three STXS bins in a phase space region with at least two jets are defined. In addition to requesting the $p_{\mathrm{T}}$ of the subleading jet to be higher than $30$ GeV and the absolute value of the Higgs rapidity ($\abs{ y_\mathrm{H}}$) less than $2.5$, the VBF STXS bins are delineated by one of the following cuts on $m_{\mathrm{jj}}$ and $p_{\mathrm{T}}^{\mathrm{H}}$:
\begin{itemize}
    \item $350 < m_{\mathrm{jj}} \leq 700$ and $p_{\mathrm{T}}^\mathrm{H} < 200 $ GeV,
    \item $ m_{\mathrm{jj}} > 700$ and $p_{\mathrm{T}}^\mathrm{H} < 200 $ GeV,
    \item $m_{\mathrm{jj}} > 350 $ GeV and $p_{\mathrm{T}}^\mathrm{H} > 200 $ GeV.
    \end{itemize}
    
Ahead of these requirements, a global selection, that mimics the analysis selection criteria at reconstructed-level, has been applied to outline a set of signal-enriched phase space regions. However, in the last two bins the number of events was found to be too low to constitute a sufficiently large training sample, and therefore it was decided to merge them together. Although they are very distant regions of phase space, it is useful to test the performance of the adversarial approach in the merged category.  \\
For the sake of simplifying the notation, the phase space where $350 < m_{\mathrm{jj}} \leq 700$, $p_{\mathrm{T}}^\mathrm{H} < 200 $ GeV  and $\abs{ y_\mathrm{H}}<2.5$ will be referred to as category 1 or simply C1, whereas the region where $ m_{\mathrm{jj}} > 700$, $p_{\mathrm{T}}^\mathrm{H} < 200 $ GeV and $\abs{ y_\mathrm{H}}<2.5$ or $m_{\mathrm{jj}} > 350 $ GeV, $p_{\mathrm{T}}^\mathrm{H} > 200 $ GeV and $\abs{ y_\mathrm{H}}<2.5$ will be named category 2 or C2. The analysis is thus performed within the particle-level categories summarized in Table~\ref{phase_space}. 
\begin{table*}[t]
    \centering
    \caption{\emph{Definition of the analysis phase space. All the variables used for the analysis requirements are defined using particle-level quantities.}}
    \begin{tabular}{c c}
    \toprule
    \multicolumn{2}{c}{Global selection} \\ \toprule
    \multicolumn{2}{c}{ oppositely-charged $\mathrm{e}\upmu$ final state, } \\
    \multicolumn{2}{c}{ at least two jets with $p_{\mathrm{T}} > 30$ GeV,}\\
    \multicolumn{2}{c}{ $p_{\mathrm{T}}^{\mathrm{\ell_1}} > 25$ GeV, $p_{\mathrm{T}}^{\ell_2} > 13$ GeV, $p_{\mathrm{T}}^{\ell_3} < 10$ GeV,} \\
    \multicolumn{2}{c}{ $m_{\ell\ell} > 12$ GeV, $p_{\mathrm{T}}^{\ell\ell} > 30$ GeV, $E_\mathrm{T}^{\mathrm{miss}} > 20$ GeV,} \\
    \multicolumn{2}{c}{ $m_\mathrm{T}^\mathrm{H} > 60$ GeV, $m_\mathrm{T}^{\ell_2} > 30$ GeV,}\\ 
    \multicolumn{2}{c}{ $\abs{\eta_{\mathrm{j}_1}} < 4.7$, $\abs{\eta_{\mathrm{j}_2}} < 4.7$ } \\
    \midrule
    C1 & C2 \\
    \midrule
    \multirow{3}{*}{$350 < m_{\mathrm{jj}} \leq 700$ GeV, $p_{\mathrm{T}}^\mathrm{H} < 200 $ GeV, $\abs{ y_\mathrm{H}}<2.5$} 
    & $ m_{\mathrm{jj}} > 700$ GeV, $p_{\mathrm{T}}^\mathrm{H} < 200 $ GeV, $\abs{ y_\mathrm{H}}<2.5$ \\
    & or \\
    & $m_{\mathrm{jj}} > 350 $ GeV, $p_{\mathrm{T}}^\mathrm{H} > 200 $ GeV, $\abs{ y_\mathrm{H}}<2.5$ \\
    \bottomrule
    \end{tabular}
    \label{phase_space}
    \end{table*}
The adopted baseline selection is inspired by the ones developed by the CMS collaboration for previous works~\cite{CMS:2022uhn}: at least two jets having $p_{\mathrm{T}}>30$ GeV each and an invariant mass higher than $120$ GeV are required; the leading and subleading leptons  ($\ell_1$ and $\ell_2$ respectively) must pass $p_{\mathrm{T}}$ thresholds dictated by the trigger requirements ($p_{\mathrm{T}}^{\ell_1}>25$ GeV and $p_{\mathrm{T}}^{\ell_2}>13$ GeV), while the $p_{\mathrm{T}}$ of the third lepton, if present, is demanded to be below $10$ GeV in order to suppress minor backgrounds, such as WZ and triboson production. Moreover, the dilepton invariant mass is required to be higher than $12$ GeV to reduce the contribution from Quantum Chromodynamics (QCD) and low mass resonances. The condition of having different flavour lepton pairs in the event suppresses Drell-Yann (DY) production of $\mathrm{ee}$ and $\upmu\upmu$ pairs, while the cuts on the dilepton transverse momentum ($p_{\mathrm{T}}^{\ell\ell}$) and the transverse mass of the system ($m_\mathrm{T}^\mathrm{H}$), defined as  \begin{equation}
    m\mathrm{^H_T} = \sqrt{2p^{\ell\ell}_\mathrm{T} E_\mathrm{T}^{\mathrm{miss}}[1-\cos\Delta\phi(\vec{p}^{\  \ell\ell}_\mathrm{T}, \vec{E}^{\  \mathrm{miss}}_\mathrm{T})]},
\end{equation} are necessary to reduce the DY production of ${\uptau\uptau}$ pair. An amount of $E_\mathrm{T}^{\mathrm{miss}}$ higher than $20$ GeV is requested because of the presence of the neutrinos in the event. The $m_\mathrm{T}^{\ell_2}$ variable, defined as
    \begin{equation}
        m_\mathrm{T}^{\ell_2} = \sqrt{2p^{\ell_2}_\mathrm{T} E_\mathrm{T}^{\mathrm{miss}}[1-\cos\Delta\phi(\vec{p}^{\  \ell_2}_\mathrm{T}, \vec{E}^{\  \mathrm{miss}}_\mathrm{T})]}
    \end{equation}
is required to be above 30 GeV, helping to reduce the non-prompt lepton background. 
Finally, the absolute value of the pseudorapidity of the leading and subleading jets ($\eta_\mathrm{j_1}$ and $\eta_\mathrm{j_2}$ respectively) have to be smaller than $4.7$, to mimic the detector acceptance. The effect of the b-tagging algorithm that is used for the identification of jets coming from b quarks in real analyses is taken into account by applying the b-tagging efficiency on MC events at particle level: a b-jet tagging efficiency $\epsilon = 94\%$ is assumed and all the MC simulated events are weighted by $(1-\epsilon)^{\mathrm{n_b}}$, where $n_b$ is the number of jets in the event that contain a B-hadron.

Two adversarial DNNs have been trained, sharing the same architecture described below, within the C1 and C2 phase spaces, respectively.\\
For what the classifier structure is concerned, three classes as targets have been defined, one for \emph{VBF} and two for the backgrounds that have to be separated from the signal: \emph{ggH} and \emph{BKG}. \\The label \emph{ggH} refers to gluon-gluon fusion (ggH) processes while the \emph{BKG} embeds both top quark events and non-resonant WW production. \\
The input variables $x_1, ..., x_m$ of this network are the measurable kinematic variables of an event, which are detailed in the following.
The activation function of the hidden layers is the rectified linear activation function (ReLU), while the output layer uses the softmax function.\\
The hyperparameters $n\mathrm{_l^C}$, $\eta^\mathrm{C}$ and $n_{\mathrm{nodes}}$, which correspond to the number of hidden layers, the learning rate and the number of nodes in the hidden layers, respectively, have been optimized through the procedure detailed below.
\\
Since six possible alternative physics model with respect to the SM have been considered, the adversary has instead seven classes, labeled \emph{VBF$\_$SM},\emph{VBF$\_$H0M}, \emph{VBF$\_$H0PH}, \emph{VBF$\_$H0L1}, \emph{VBF$\_$H0Mf05}, \emph{VBF$\_$H0PHf05}, and \emph{VBF$\_$H0L1f05}, respectively. The first string refers to the VBF signal events generated assuming the SM hypothesis; the second one labels the signal events obtained  according to H0M model, and so on.\\
The input layer of this network is the second-to-last layer of the classifier. The ReLU activation function is used for the hidden layers, while the softmax function is used for the output layer.\\
As for the classifier, the number of hidden layers $n\mathrm{_l^A}$, the learning rate $\eta^\mathrm{A}$ and the number of nodes in each hidden layer $n_{\mathrm{nodes}}$ have been optimized.
Note that the same number of nodes in the hidden layers for both $C$ and $A$ has been chosen.

\subsection{Training procedure}
The training has been carried out on MC simulations and the following samples have been chosen: SM ggH production, SM non-resonant WW and top quark production, which constitute the background event samples, and finally the VBF signal. The latter has been generated separately according to each of the seven considered physical hypotheses.

MC samples are simulated at various perturbative orders in perturbative-QCD (pQCD), using different event generators.\\ 
Gluon-gluon fusion and vector boson fusion processes are generated with \textsc{powheg} v2 \cite{Nason:2004rx} at next-to-leading order (NLO) accuracy in QCD. The subsequent decay of the Higgs boson in two W bosons is performed using \textsc{jhugen} v7.1.4, which also simulates the leptonic decay of the vector bosons. The BSM samples which describe anomalous HVV couplings are generated with \textsc{powheg} v2.
For what concerns background processes, the non-resonant WW events sample is simulated using two different generators, depending on the production processes: $\mathrm{q\bar{q}} \rightarrow \mathrm{WW}$ is generated with \textsc{powheg} v2 at NLO accuracy while the $\mathrm{gg }\rightarrow \mathrm{WW}$ events simulation is performed by \textsc{mcfm} v7.0.1 at leading order (LO) accuracy. Single top and top-antitop pair production are generated with \textsc{powheg} v2 at NLO accuracy.
In order to provide simulation of initial and final state radiation, hadronization, and underlying event, all the event generators are interfaced to \textsc{pythia} 8.1 \cite{Sjostrand:2007gs}.

The events entering the training are selected through the kinematic requirements corresponding to the global selection listed in Table~\ref{phase_space}.
\newline
The training procedure has been performed using the Keras~\cite{chollet2015keras} and TensorFlow~\cite{abadi2016tensorflow} libraries.

A set of input variables that highlight the signal characteristics with different degrees of discriminating power with respect to the other processes has been defined. The input variables chosen for the training are defined at particle-level and are listed below: 
\begin{itemize}

    \item $p\mathrm{_{T_{j_1}}}$ , $p\mathrm{_{T_{j_2}}}$: the magnitudes of the transverse momenta of the leading and subleading jet, respectively;

    \item  $\eta_{\mathrm{j}_1}$: the pseudorapidity of the leading jet $j_1$. Since it is typically generated by the hadronization of a $q$ quark with a high fraction of the proton four-momentum, $j_1$ coming from the VBF process tends to be emitted at large $\abs{\eta}$. Instead, background events are characterized by the presence of a leading jet in the central region i.e., at low $\abs{\eta}$;
   
    \item $\eta_{\mathrm{j}_2}$: the pseudorapidity of the subleading jet. Its distribution is similar to $\eta_{j_1}$ for the VBF process, while background events show homogeneous shapes throughout the range of possible values of this variable;

    \item  $\abs{\Delta\eta_{\mathrm{jj}}}$: the separation in $\eta$ between the two jets in the final state. The signal shows a larger pseudorapidity gap with respect to the backgrounds;

    \item $m_{\mathrm{jj}}$: the invariant mass of the dijet system. It has a good discrimination power since it is typically larger for the signal than for the background;
    
    \item $p_{\mathrm{T}}^{\ell\ell}$, $p_{\mathrm{T}}^{\ell_1}$, $p_{\mathrm{T}}^{\ell_2}$: the magnitudes of the transverse momenta of the dilepton system, the leading lepton, and the subleading lepton, respectively;
    
    \item $\eta_{\ell_1}$, $\eta_{\ell_2}$: the pseudorapidity of the leading and subleading lepton, respectively. Similarly to $p_{\mathrm{T}}^{\ell\ell}$, $p_{\mathrm{T}}^{\ell_1}$ and $p_{\mathrm{T}}^{\ell_2}$, these variables are not expected to have a high discrimination capability individually, but are included to exploit their correlation with other observables;

    \item $m_{\ell\ell}$: the invariant mass of the lepton pair. Due to the spin correlation effect in the $\mathrm{H\rightarrow WW\rightarrow2\ell2\nu}$ decay chain, this variable is peaked at low values for the VBF and ggH mechanisms, while showing a broadened shape for non-resonant events;
    
    \item $\Delta\phi_{\ell\ell}$: the angular separation in $\phi$ between the two leptons in the final state. For signal events, it has a smaller value than in the non-resonant case, since the spin correlation effect forces the directions of the two charged leptons to be nearly collimated;
    
    \item $\Delta R_{\ell\ell}$: the radial separation between the two leptons in the final state. Similarly to $\Delta\phi_{\ell\ell}$, this variable is almost flat for events without a resonance;

    \item  $m_{\ell \mathrm{j}}$: the invariant mass of the system consisting of the $\ell$-th lepton and $j$-th jet, where $\ell=\{\ell_1, \ \ell_2\}$ and $j=\{j_1, \ j_2\}$. There are four possible combinations which have limited, but not completely negligible, discrimination power;
    
    \item $C_{\mathrm{tot}}=\log\Bigl(\sum\limits_{\ell}\abs{(2\eta_{\ell}-\sum\limits_{\mathrm{j}}\eta_\mathrm{j})}/\abs{\Delta\eta_{\mathrm{jj}}}\Bigr)$, where $\ell=\{\ell_1, \ \ell_2\}$, $j=\{j_1, \ j_2\}$: it represents a measure of how much the charged leptons are emitted centrally with respect to the dijet system;
    
    \item $E_\mathrm{T}^{\mathrm{miss}}$: the missing transverse energy;
    
    \item $m\mathrm{^H_T}$: the transverse mass of the system. Background events are more likely to present a small value of this variable;
    
    \item $m\mathrm{^{vis}}=\sqrt{(p^{\ell\ell} +E_\mathrm{T}^{\mathrm{miss}})^2-(\vec{p}^{\ \ell \ell }+\vec{E}^{\ \mathrm{miss}}_\mathrm{T})^2}$: the visible mass, which includes the longitudinal momentum of the dilepton system. It tends to have a different shape for events containing the production of a Higgs boson;
    
    \item $\Delta\phi(\vec{p}^{\ \ell\ell}_\mathrm{T}, \vec{E}^{\ \mathrm{miss}}_\mathrm{T})$: the azimuthal opening angle between $\vec{p}^{\ \ell\ell}_\mathrm{T}$ and $\vec{E}^{\ \mathrm{miss}}_\mathrm{T}$. It has a good discriminating power and it is included in the definition of $m\mathrm{^H_T}$;

    \item $H_\mathrm{T}$: the scalar sum of the transverse momenta of all jets in the event. It gives a measure of the hadronic activity of the event.
    
\end{itemize}

\subsection{Optimization and performance}\label{sec:opt}
As anticipated before, since the kinematic properties of signal and background events depend on the values of $m_{\mathrm{jj}}$ and $p_{\mathrm{T}}^\mathrm{H}$, in order to add more flexibility to the model and test it within two different phase space regions, two different ADNNs have been implemented: the first network is trained on events with \begin{equation}\label{bin1}
    350 < m_{\mathrm{jj}} \leq 700 \textnormal{ GeV, } p_{\mathrm{T}}^\mathrm{H} < 200  \textnormal{ GeV and }  \abs{ y_\mathrm{H}}<2.5 ,
\end{equation} 
while the second one learn on events with\begin{equation}\label{bin2}
    m_{\mathrm{jj}} > 700 \textnormal{ GeV, } p_{\mathrm{T}}^\mathrm{H} < 200  \textnormal{ GeV and  } \abs{ y_\mathrm{H}}<2.5 \textnormal{ or }  m_{\mathrm{jj}} > 350  \textnormal{ GeV, } p_{\mathrm{T}}^\mathrm{H} > 200  \textnormal{ GeV and } \abs{ y_\mathrm{H}}<2.5,
\end{equation}
namely the phase space region C1 and C2 defined in the previous Section.
Therefore, requirements~\ref{bin1} and~\ref{bin2} are added to selection criteria of training events of the first and the second ADNN respectively. The decision to train two different ADNNs in C1 and C2 is not strictly necessary, but it allows testing the capability of the adversarial approach on more than one phase space region.

The training samples of the ADNN for~\ref{bin1} is composed by about $65000$ events, while the ADNN~\ref{bin2} is trained on about $47000$ events. Both training samples are equally divided between the three considered physical processes, i.e. VBF, ggH and background events. The VBF signal sample is in turn made up of events coming from the seven possible domains in equal proportions. 
The sample of background events is made of non-resonant WW (including the gluon-induced process and the electroweak production), $\mathrm{t}\bar{\mathrm{t}}$ and $\mathrm{tW}$ events, according to the proportions predicted by the SM.\\ 
Both the training samples have been divided into two subsets: 80$\%$ of the total events is used for the learning procedure whereas the remaining 20$\%$ is reserved for validation.

The hyperparameters of the ADNNs have been optimized following the procedure explained in Section~\ref{sec:ADNN_implementation}. The hyperparameters that were optimized are $n\mathrm{_l^C}$, $\eta^\mathrm{C}$, $n\mathrm{_l^A}$, $\eta\mathrm{^A}$, $n_{\mathrm{nodes}}$ and $\alpha$ for both ADNNs; for the optimization, the Optuna software~\cite{optuna_2019} was employed, which enabled the maximization of the classifier categorical accuracy and minimization of the average K-S test statistic simultaneously. Both the chosen objective functions have been evaluated on the validation set at the end of each optimization trial. 
First, 100 training trials with 800 epochs each have been executed, varying the values of all the hyperparameters according to a Bayesian optimization approach~\cite{pmlr-v54-klein17a} within defined ranges. Then, the training which provided the best combination of a high categorical accuracy value and a low average K-S test statistic among all the attempts has been chosen as best trial and the $n\mathrm{_l^C}$, $n\mathrm{_l^A}$ and $n_{\mathrm{nodes}}$ parameters have been fixed to their best value. Finally, a second set of training trials has been repeated in order to optimize the learning rates and the $\alpha$ parameter.\\ The intervals associated with possible outcomes of the hyperparameters are reported in the Table~\ref{hyp_ranges} while the hyperparameters numerical values, as well as the categorical accuracy and average K-S test values, corresponding to the chosen best trial for both the ADNNs are summarized in Table~\ref{optimization_results}. The p-value of the K-S test statistic of the ADNN trained in C1 (C2) ranges from 6$\%$ (7$\%$), for the model with the worst compatibility with the SM, to 50$\%$ (59$\%$) for the best, with a median value across the six models of 34$\%$ (41$\%$).
\begin{table}[ht]
    \centering
    \caption{\emph{Ranges within the optimization procedure can vary the hyperparameters values}.}
    \begin{tabular}{cc}
    \toprule
    Hyperparameter & Interval\\ \midrule
    $\alpha$ & $\left[0,100\right]$\\
    $\eta^\mathrm{C}$ & $\left[10^{-5},10^{-3}\right]$\\
    $\eta^\mathrm{A}$ & $\left[10^{-4},10^{-2}\right]$\\
    $n_{\mathrm{nodes}}$ & $\left[10,100\right]$\\
    $n\mathrm{_l^C}$ & $\left[1,10\right]$\\
    $n\mathrm{_l^A}$ & $\left[1,10\right]$\\
    \bottomrule
    \end{tabular}
    \label{hyp_ranges}
    \end{table}
\begin{table}[ht]
    \centering
    \caption{\emph{Categorical accuracy, average K-S test statistic and hyperparameters numerical values associated to the chosen working point of the two ADNNs.}}
    \begin{tabular}{ccc}
    \toprule
    Hyperpatameter & C1 & C2\\ \midrule
    $\alpha$ & $100$  & $68$\\
    $\eta^\mathrm{C}$ & $0.00098$ & $0.00014$\\
    $\eta^\mathrm{A}$ & $0.006$ & $0.00028$\\
    $n_{\mathrm{nodes}}$ & $48$ & $95$\\
    $n\mathrm{_l^C}$ & $4$ & $7$ \\
    $n\mathrm{_l^A}$ & $9$ & $9$\\ \midrule
    \multicolumn{3}{c}{} \\ \midrule
    Objective function & & \\ \midrule
    categorical accuracy & $69\%$ & $72\%$ \\
    average K-S test statistic & $0.05$ & $0.07$\\
    \bottomrule
    \end{tabular}
    \label{optimization_results}
    \end{table}
    
\begin{figure}[ht]
    \centering
    {\includegraphics*[width=.45\columnwidth]{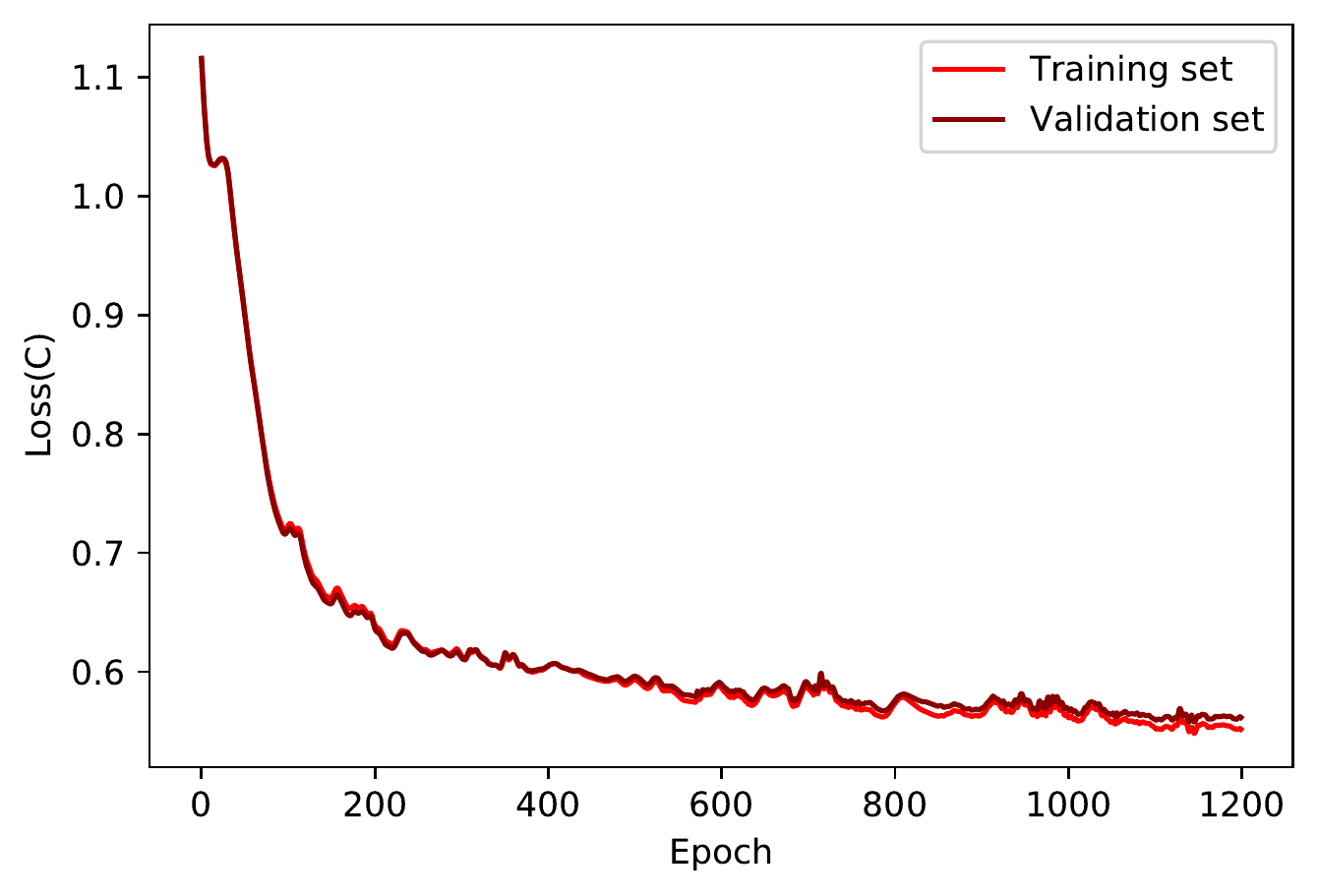}} 
    \\
    {\includegraphics*[width=.45\columnwidth]{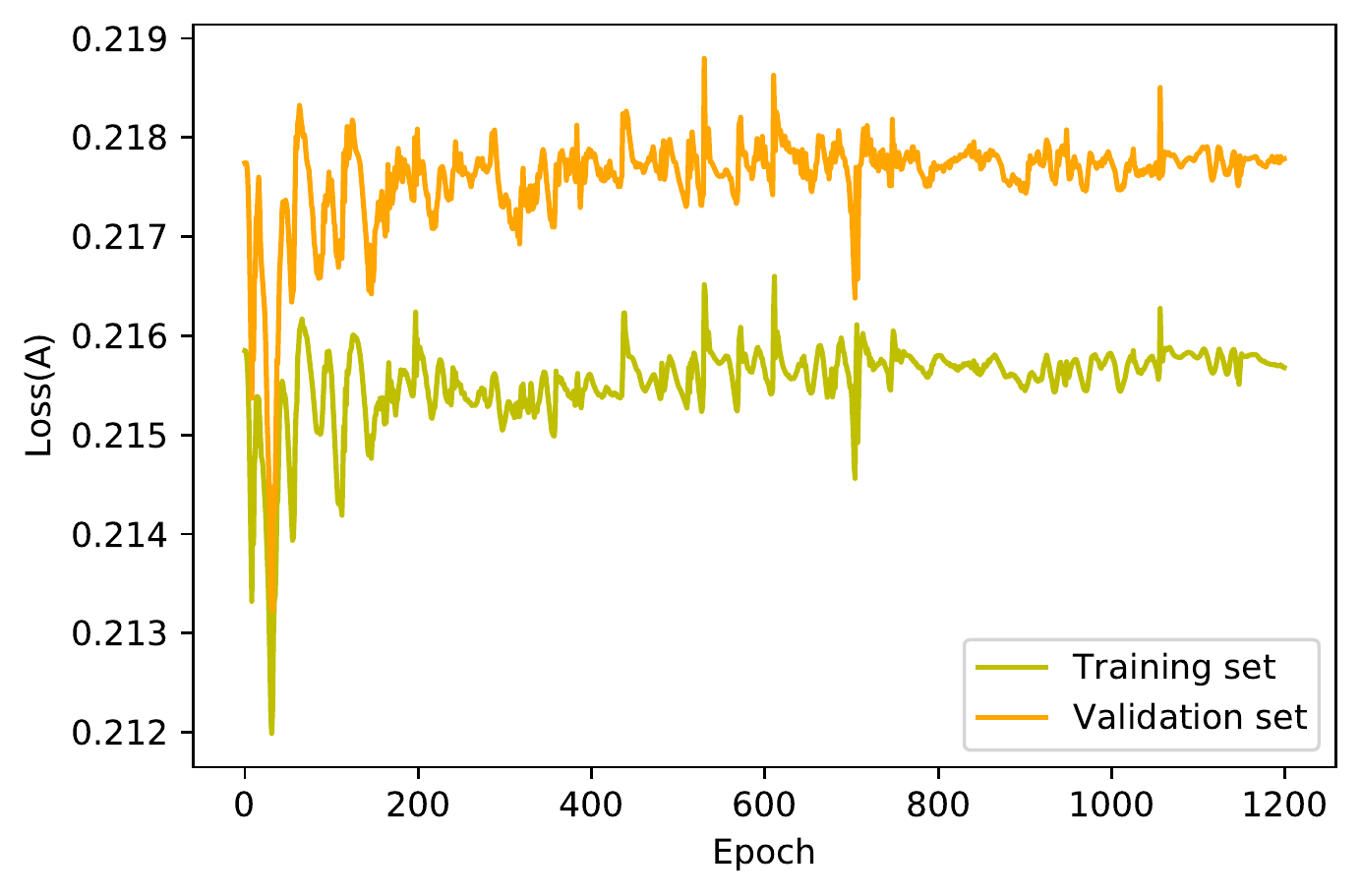}} 
    \\
    {\includegraphics*[width=.45\columnwidth]{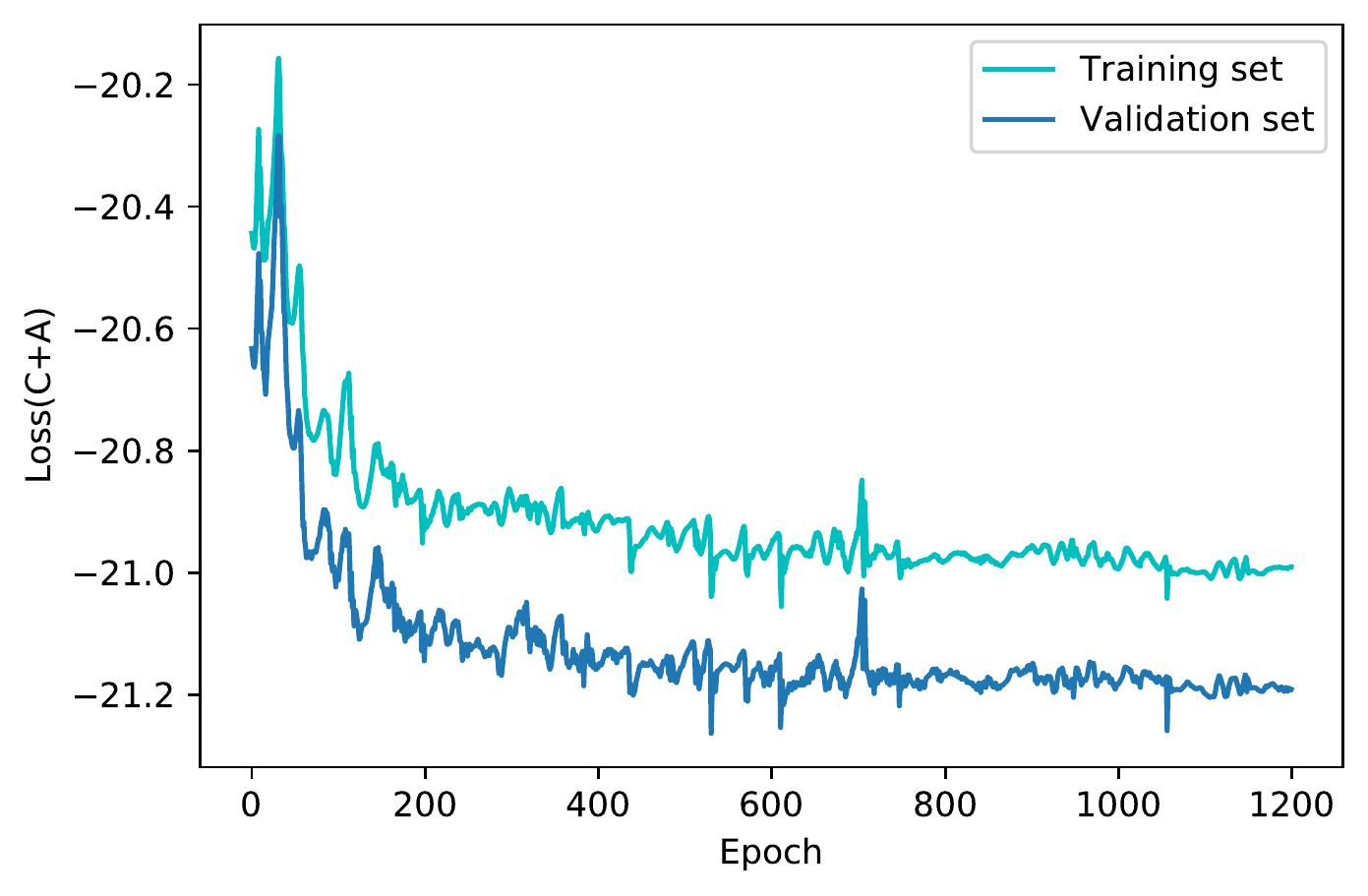}} 
    \caption{\emph{Loss function of the classifier (top), adversary (middle) and combined loss function (bottom) of the ADNN trained in C1.  }}
    \label{Losses_1}
    \end{figure}

\begin{figure}[t]
    \centering
    {\includegraphics*[width=0.45\columnwidth]{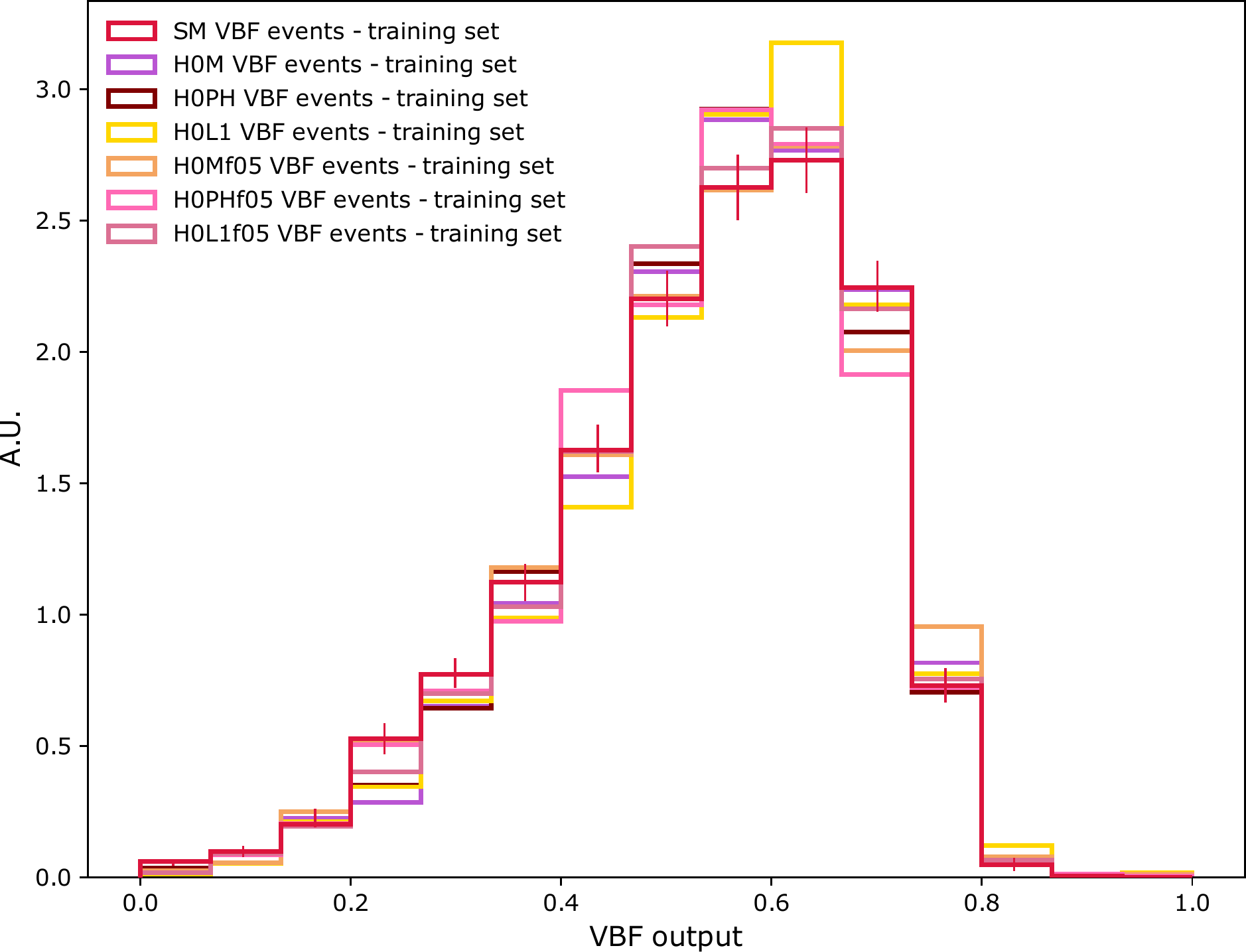} }
    {\includegraphics*[width=0.45\columnwidth]{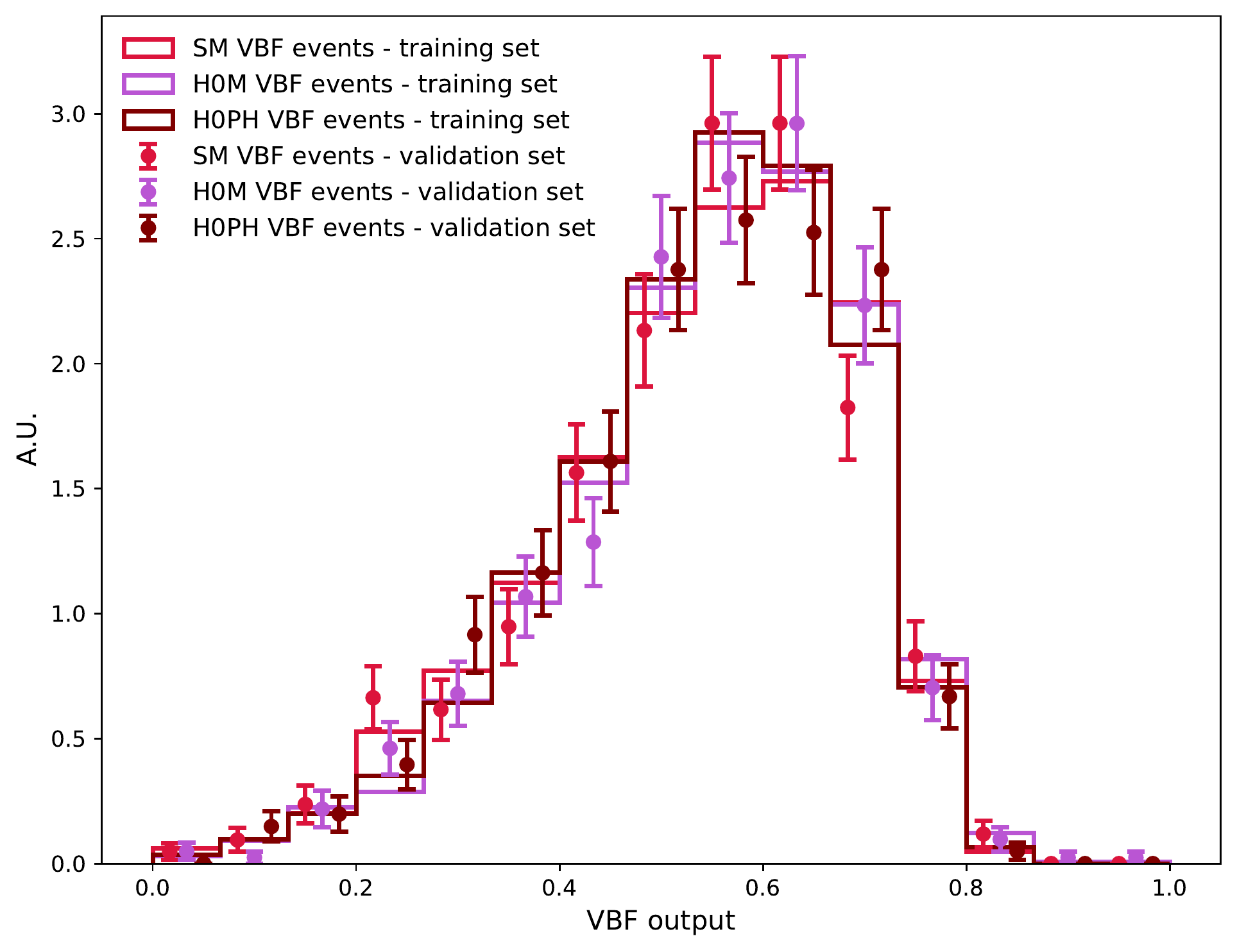} }
    \caption{\emph{Normalized distributions of the VBF-output on signal events simulated under the seven considered hypotheses, evaluated on the training sample (left) in C1. The uncertainty associated with the limited size of simulated samples is reported only in the SM distribution. Three distributions related to the SM, H0M and H0PH models have been tested simultaneously in the training and validation sample (right). The solid lines correspond to the distributions obtained on training set, while the dots with statistical error bars stand for validation sample.}}
    \label{VBF_output_1}
    \end{figure}

\begin{figure}[ht]
    \centering
    {\includegraphics[width=0.45\columnwidth]{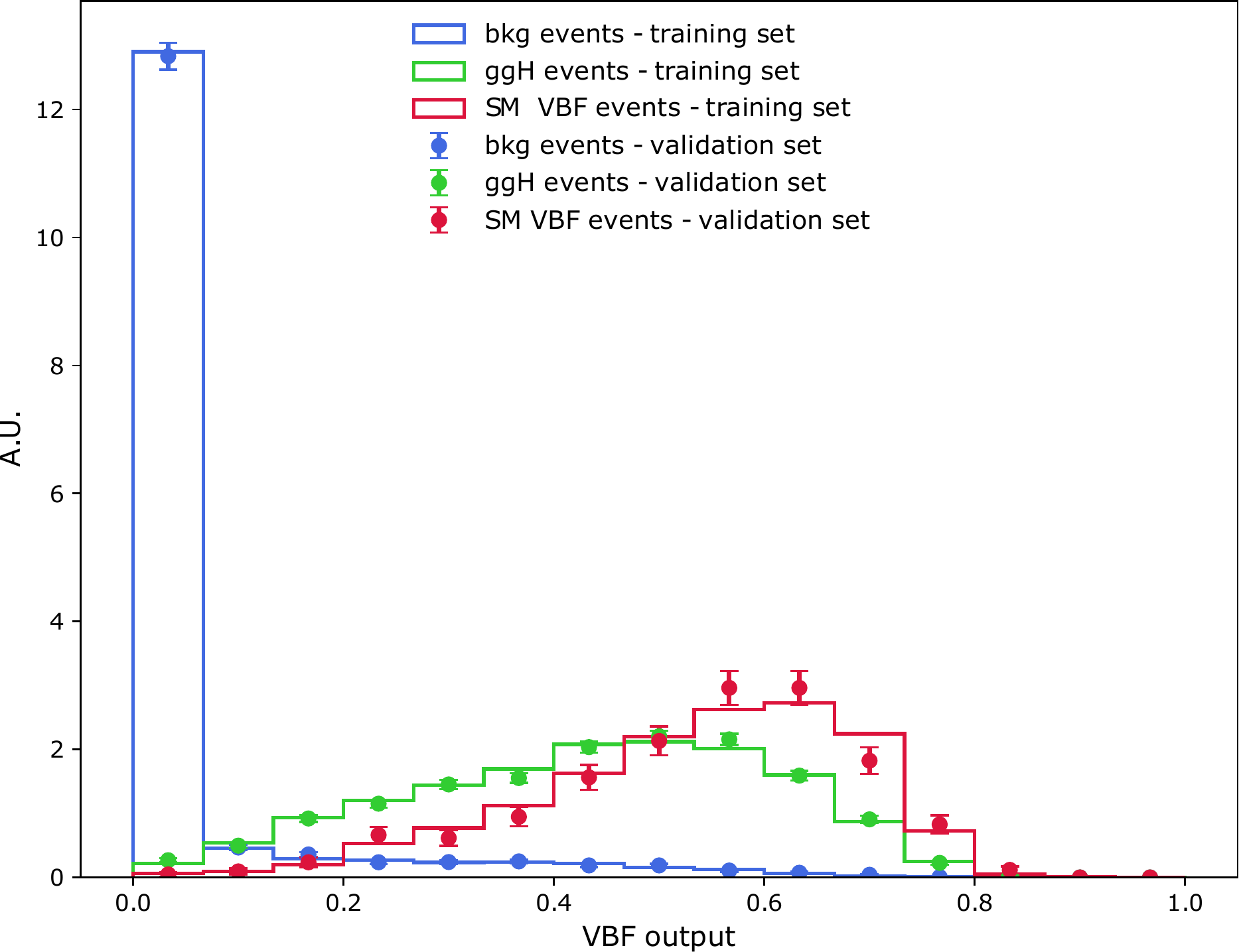}} 
    \caption{\emph{Normalized distributions of the VBF-output on SM signal, ggH  and background events in C1. The solid lines correspond to the distributions obtained on training sample, while the dots with statistical error bars stand for validation sample.}}
    \label{VBF_output_bkg_1}
    \end{figure}

\begin{figure}[ht]
    \centering
    {\includegraphics[width=0.45\columnwidth]{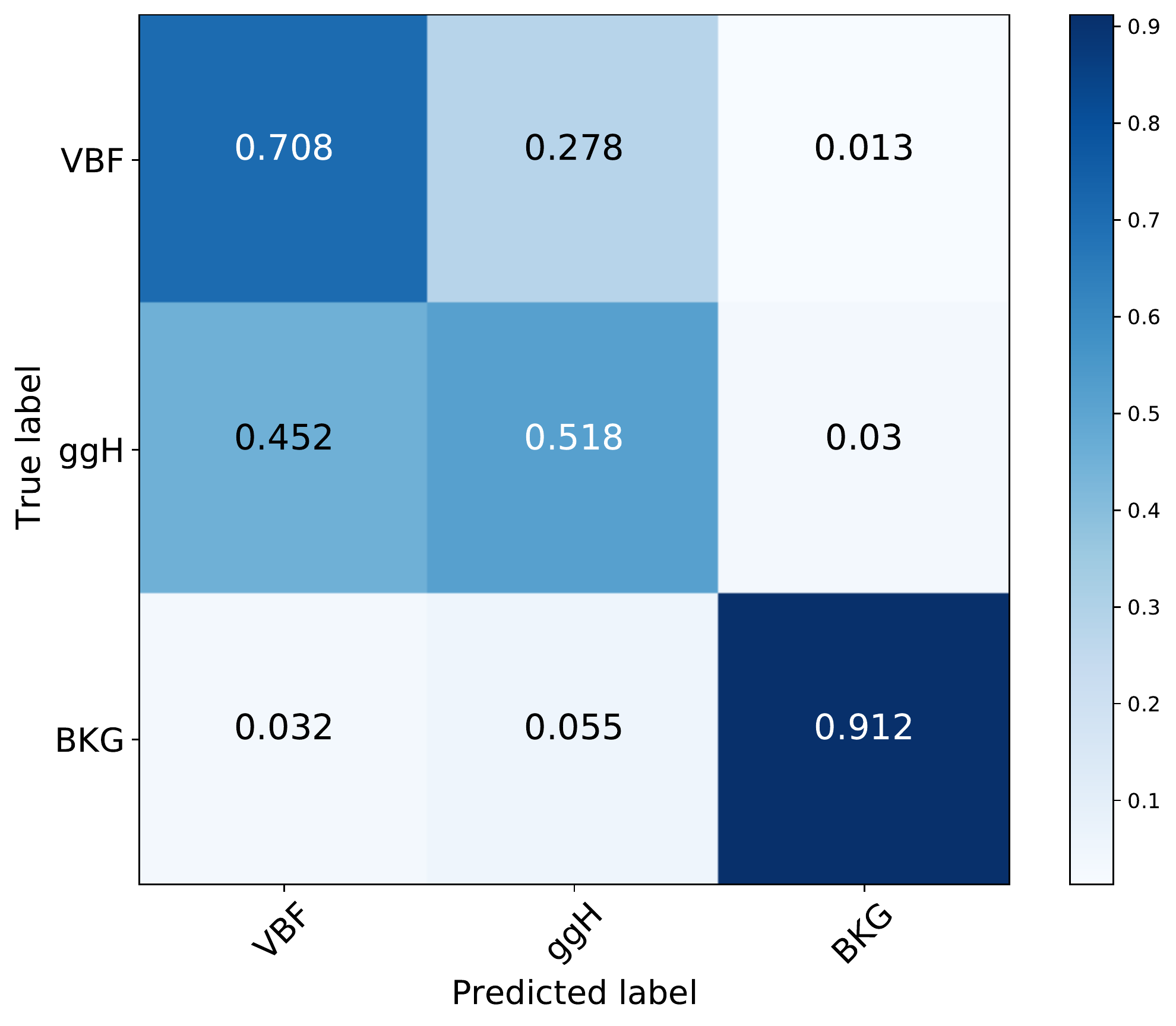}} 
    \caption{\emph{Confusion matrix of the ADNN trained in C1.}}
    \label{CM_1}
    \end{figure}

\begin{figure}[ht]
    \centering
    \subfloat[C1]
    {\includegraphics[width=.45\columnwidth]{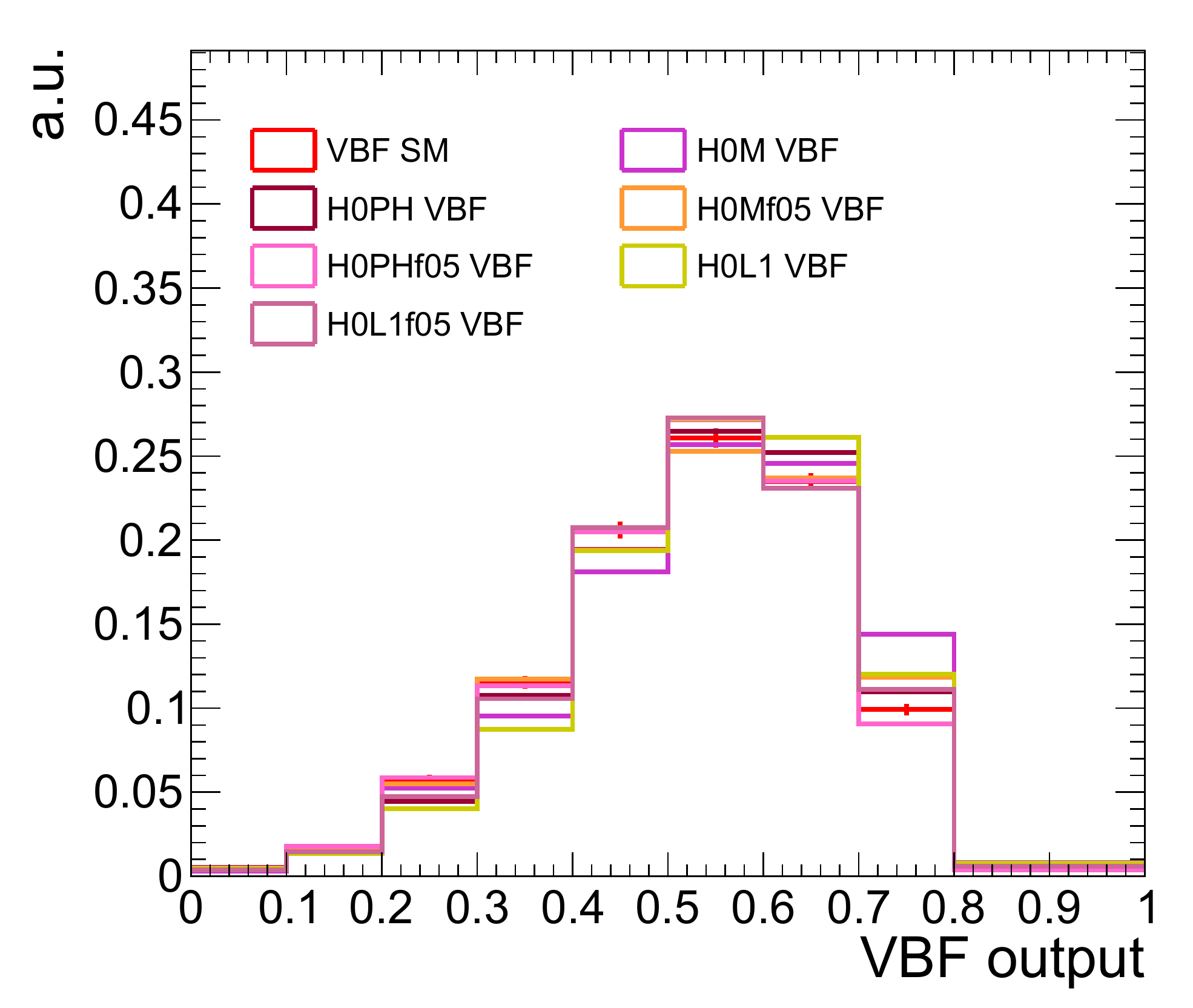}}
    \subfloat[C2]
    {\includegraphics[width=.45\columnwidth]{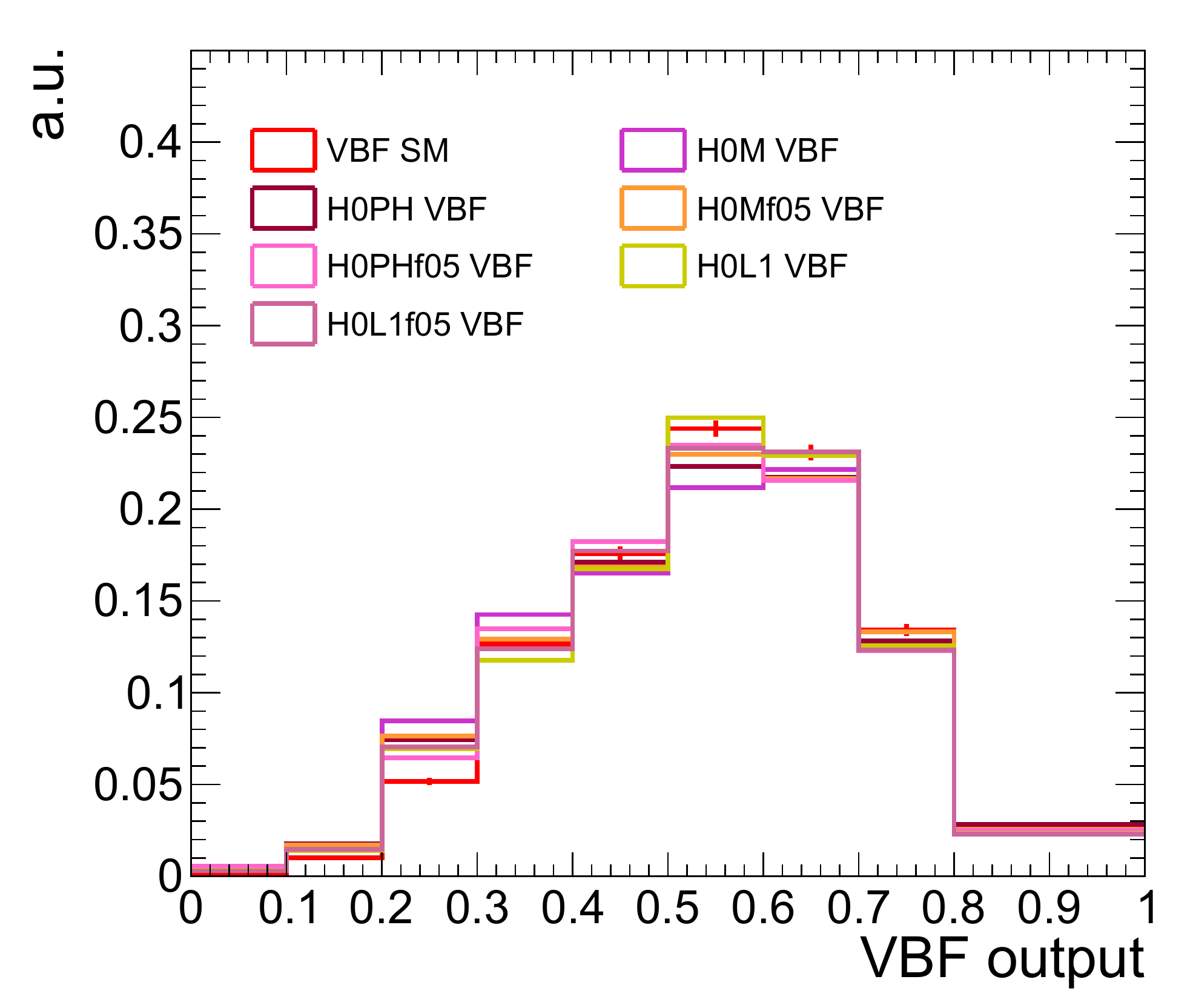}} \\
    \caption{\emph{Normalized distribution of the VBF-output of both the ADNNs in the corresponding STXS bin. The uncertainty associated with the limited size of simulated samples is reported only in the SM distribution.}}
    \label{norm}
    \end{figure}

\begin{figure}[ht]
    \centering
    \subfloat[C1]
    {\includegraphics[width=.5\columnwidth]{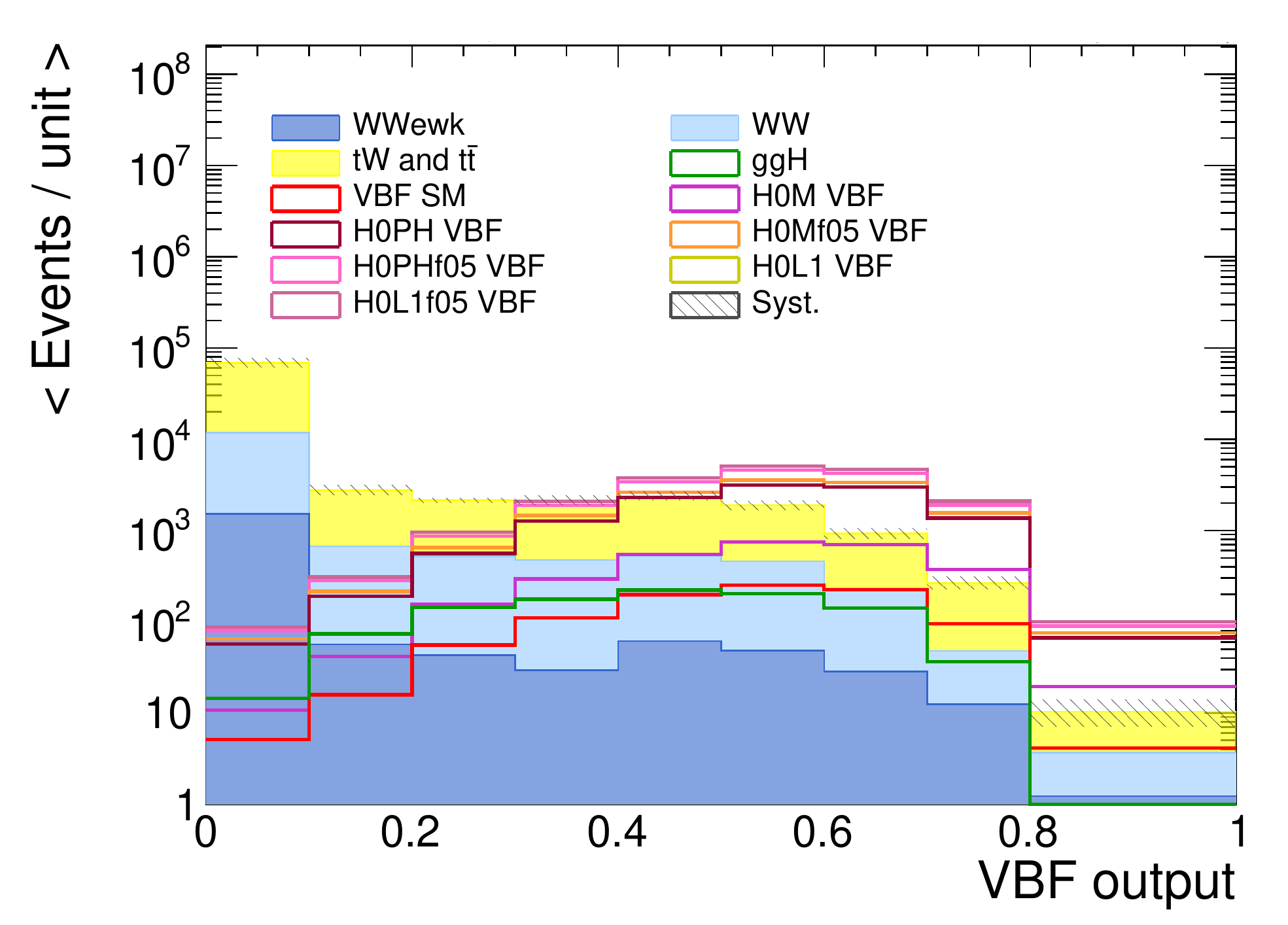}}
    \subfloat[C2]
    {\includegraphics[width=.5\columnwidth]{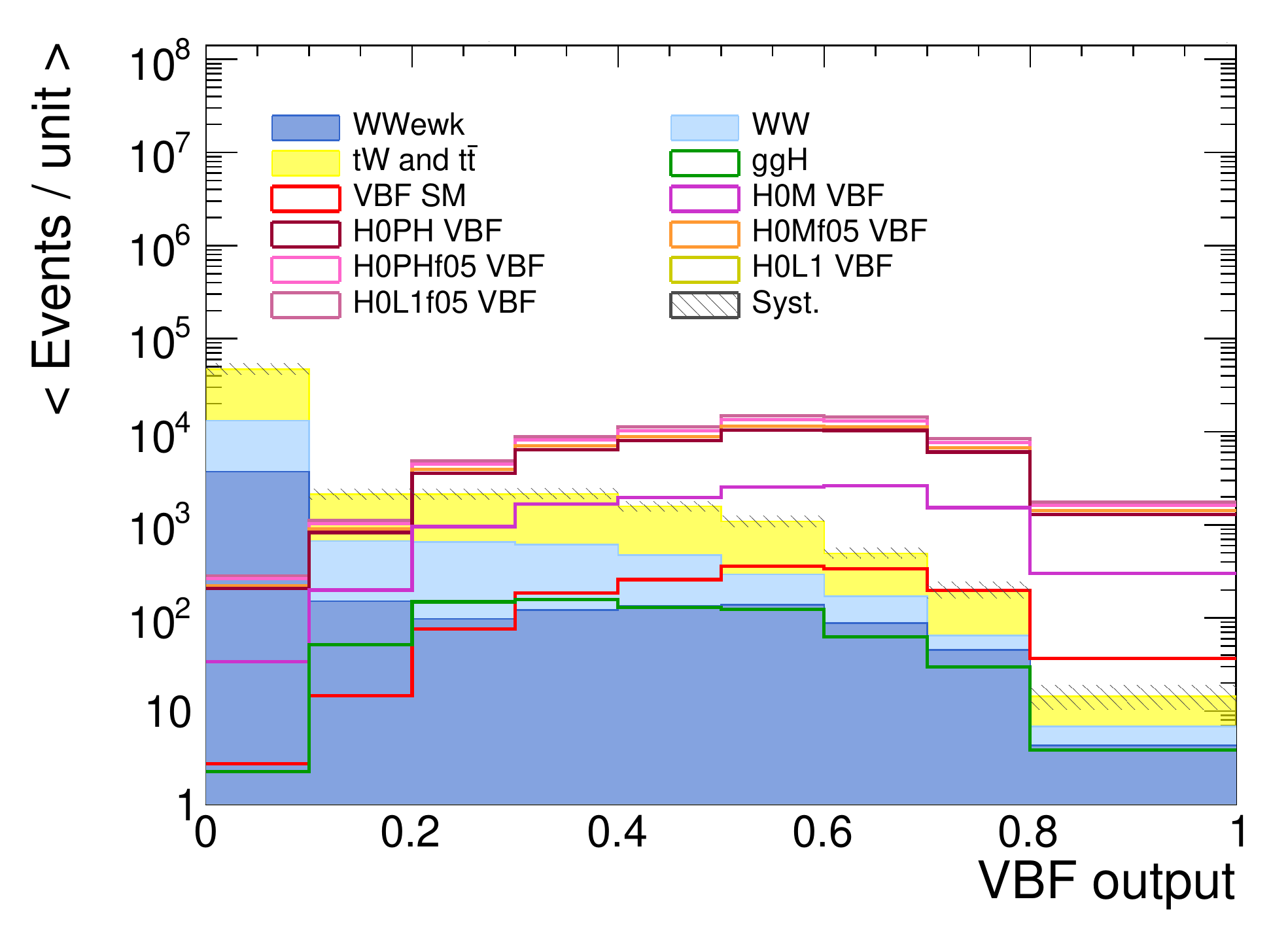}} \\
    \caption{\emph{Distribution of the VBF-output of both the ADNNs in the corresponding STXS bin. The VBF signals are shown superimposed on backgrounds templates, which are instead drawn stacked on top of each other. A logarithmic scale is used. The dashed uncertainty band corresponds to the total systematic uncertainty in the background templates.}}
    \label{sig_bac}
    \end{figure}

Once the hyperparameters have been set to their best estimation, the networks have been retrained increasing the number of learning epochs to 1200 and 1400 for the first and the second ADNNs, respectively, and with a batch size equal to the entire training sample.\\
As an example, in the following some results regarding the performances of the ADNN trained in C1 are reported. The same set of results has been studied for the second ADNN, drawing the same conclusions reported below.\\ 
Figure~\ref{Losses_1} shows the loss function of the adversary, of the classifier and the combined loss function as a function of the number of epochs. The constant trend exhibited by the loss of the adversary at the end of the training is expected, and is due to the performance of the adversary being equivalent to a random guessing.\\
In order to check the effectiveness of the procedure, the distributions of the VBF-output for signal events coming from all domains have been compared separately for each of the two ADNNs. The results are shown in Figure~\ref{VBF_output_1}. For the sake of clarity, the uncertainty associated with the limited size of simulated samples has been drawn only for the SM VBF distribution. Similar uncertainties are present also in the BSM distributions, since the number of events for each signal hypothesis in the training sample is the same.  The shapes of the distributions are in fair agreement with each other, confirming that the classifier is unable to recognize the domain of origin i.e., the signal model of the VBF events. The impact of the remaining shape differences is addressed in the bias studies which will be presented in Section~\ref{sec:bias_estimation}. Moreover, in order to check for overtraining of the algorithm, the distributions obtained from the validation sample have been compared to the ones obtained from the training one.  For the sake of clarity the distributions of only three signal models (SM, H0M and H0PH) are reported. Given that the shapes of the training distributions are compatible with the validation ones within the statistical accuracy, it was concluded that the algorithm does not show signs of overtraining.\\ 
The signal-to-background discriminating power of the ADNN is highlighted in Figure~\ref{VBF_output_bkg_1}, where only the SM VBF contribution is reported. The VBF-output exhibits a good capability of discriminating between background and signal events, whereas it is not as effective in distinguishing the VBF production mode from ggH. The performances on the validation sample are also taken into consideration and are shown superimposed to the training ones using dots with statistical error bars. \\
To assign events to classes, the class with the highest score is picked.\\
The classification accuracy of the ADNN is quantified by the confusion matrix, reported in Figure~\ref{CM_1}. The matrix shows that 71$\%$, 52$\%$ and 91$\%$ of the VBF, ggH, and background events, respectively, are correctly categorized into the corresponding class. The ADNN trained in C2 has approximately the same overall discriminating power, properly classifying about 65$\%$ events of both VBF and ggH and 90$\%$ of background.\\
Once the networks have been trained and the models parameters fixed to their best values, the ADNNs have been tested on the events selected by the analysis requirements. Figure~\ref{norm} shows the normalized distributions of the VBF-output of both ADNNs evaluated on all the considered signal models, while the distributions on signal and background events are reported in Figure~\ref{sig_bac}, evaluating each ADNN in the corresponding STXS bin. In both figures the contributions of the various processes that are expected assuming a total integrated luminosity equal to 138 fb$^{-1}$ are illustrated. The following binning scheme is adopted: $[0.0, 0.1, 0.2, 0.3, 0.4, 0.5, 0.6, 0.7, 0.8, 1.0]$.

\begin{figure}[ht]
    \centering
    \subfloat[C1]
    {\includegraphics[width=.45\columnwidth]{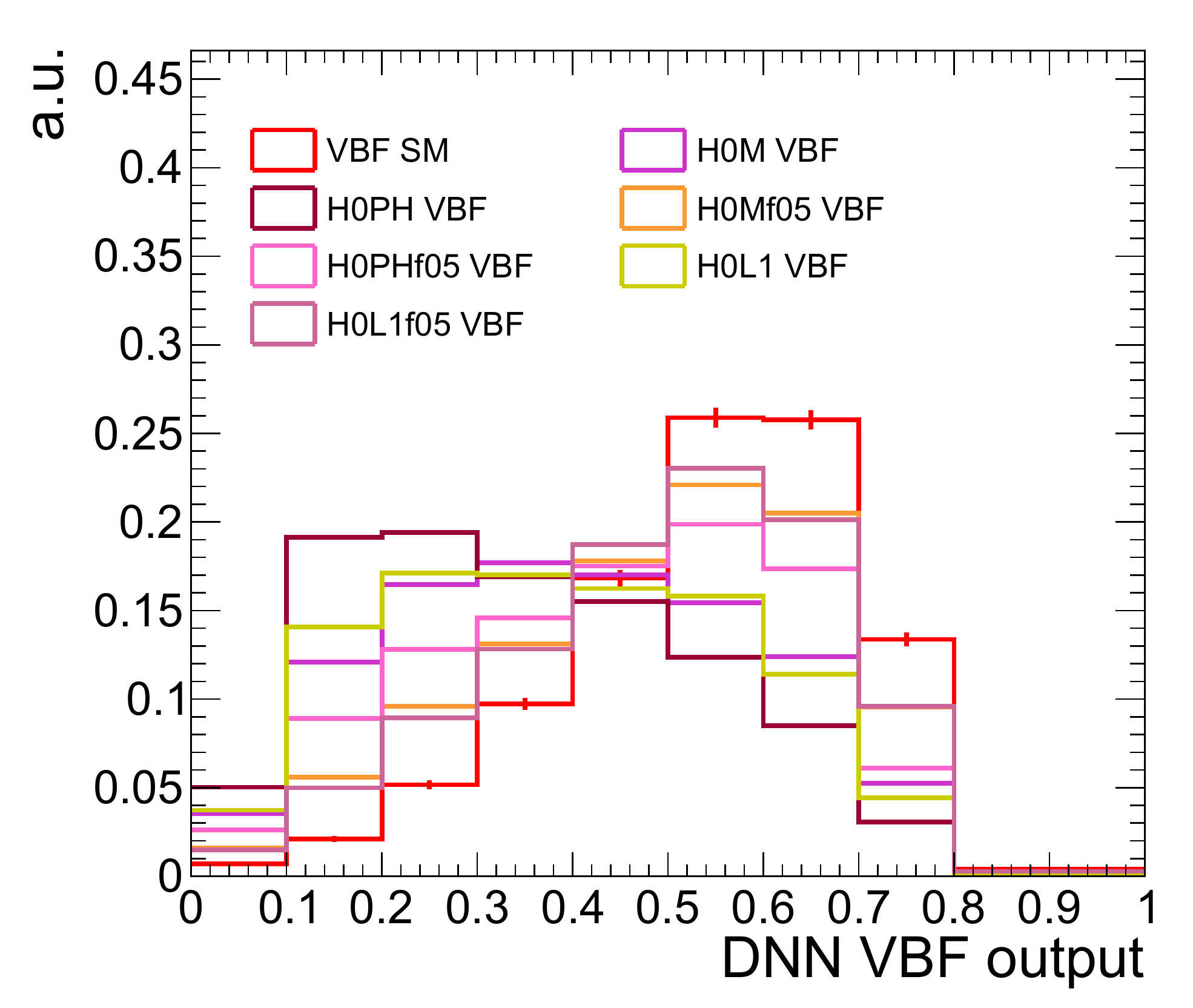}}
    \subfloat[C2]
    {\includegraphics[width=.45\columnwidth]{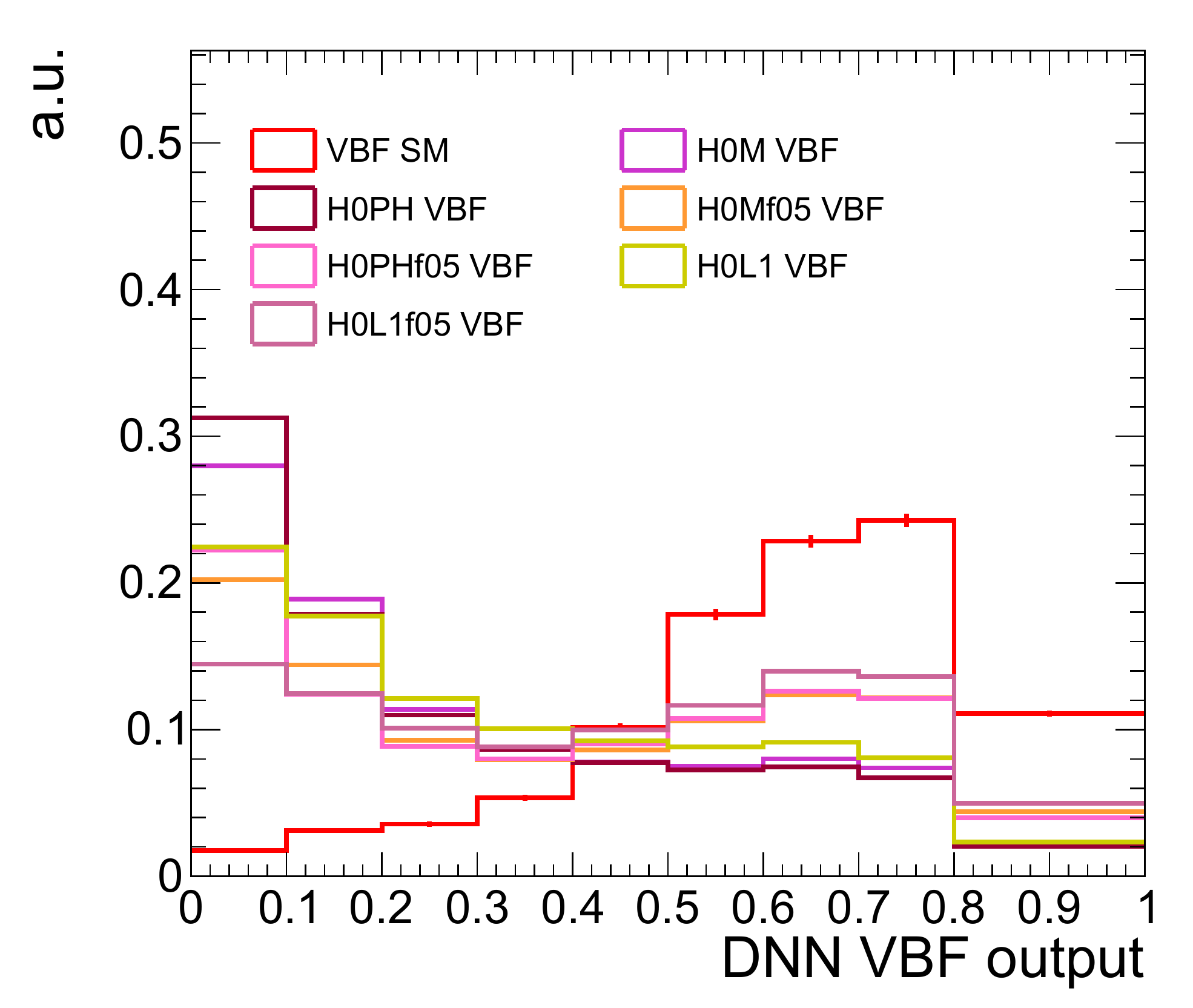}} \\
    \caption{\emph{Normalized distribution of the VBF-output of both the DNNs in the corresponding STXS bin. The uncertainty associated with the limited size of simulated samples is reported only in the SM distribution.}}
    \label{DNN_VBF_output}
    \end{figure}

With the aim of comparing the effectiveness of the ADNN with a standard feed-forward DNN used for discriminating purposes, a simpler DNN without the adversarial term and only trained on the SM signal has been also implemented. The entire analysis procedure has been repeated employing for the event categorization and for the template fit two DNNs, one for each of the considered bins. These networks take the same kinematic distributions as input variables and their learning rate, number of hidden layers and number of nodes in each hidden layer were optimized by maximizing the categorical accuracy. Similarly to what has been done for the optimization of the ADNNs, 100 training trial of 300 epochs each has been generated, varying the hyperparameters according to the Bayesian approach within the same ranges reported in Table~\ref{hyp_ranges}.\\
The best trial chosen for the DNN trained in C1 provides a categorical accuracy of 72$\%$ and requires only one hidden layer, composed of 60 nodes and a learning rate of $0.00055$. Instead, the DNN trained in C2 is composed of one hidden layers of 86 nodes, has a learning rate o $0.00079$ and a categorical accuracy of $78\%$.\\
Since the adversarial term is missing, the DNNs performance is expected to strongly depend on the signal physics model assumption. This behavior is indeed observed in Figure~\ref{DNN_VBF_output}, where the normalized distributions of the VBF-output for SM signal events and events stemming from the alternative signal models are shown. The different shapes are due to the presence of input features that have distributions strongly dependent on the theoretical signal modeling, such as $m_T^H$, $\Delta\eta_{jj}$, $H_T$ and $p_T$ of the jets.

\begin{figure}[ht]
    \centering
    {\includegraphics[width=.45\columnwidth]{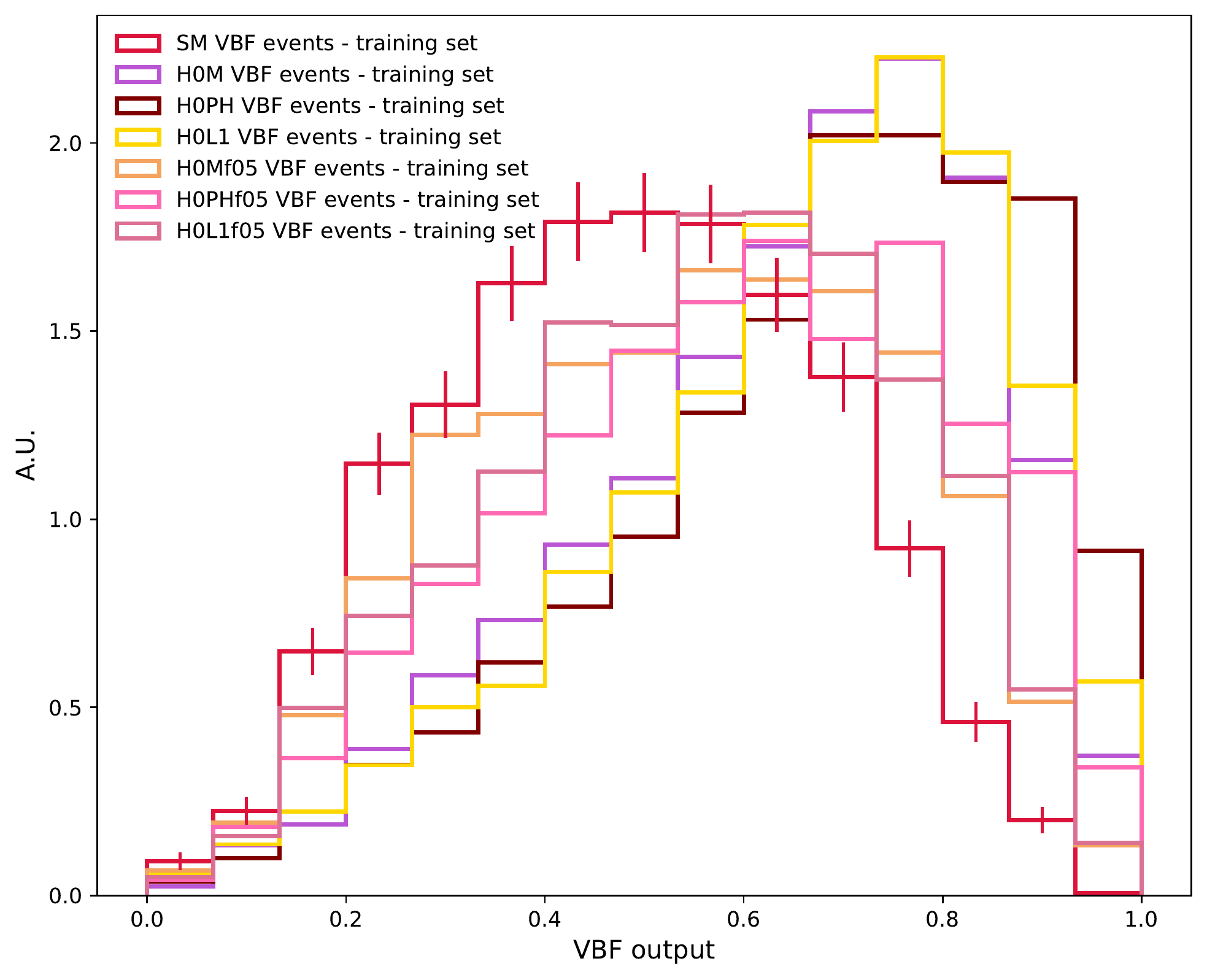}}
    \caption{\emph{Normalized distribution of the VBF-output of the DNN trained with both the SM and BSM signal events in C1. The uncertainty associated with the limited size of simulated samples is reported only in the SM distribution.}}
    \label{DNN_allBSM}
    \end{figure}

As a further study, a DNN has been trained on a VBF signal sample constructed with a mixture of SM and BSM models in equal proportions, together with the SM gluon-gluon fusion and background events. This approach can be seen as intermediate in complexity between the DNN trained on the SM only and the ADNN. The structure of this DNN has been defined by following the same optimization strategy described for the DNN trained on SM events only. For the case of the C1 category, for example, an architecture  composed of 2 hidden layers of 77 nodes each was defined and trained with 250 epochs and a learning rate equals to $0.00077$. The distributions of the VBF-output on both SM and BSM signal events are shown in Figure~\ref{DNN_allBSM}: shapes differ significantly, albeit less than in the case of the SM-trained DNN, showing a residual degree of dependence on the physics model assumption for the signal process. In particular, the network is able to better discriminate the BSM signal with respect to the SM hypothesis. Therefore, even a DNN trained on a mixture of different signal hypotheses is not able to provide a model independent classification at a level that is comparable to that achieved with the ADNN.

\subsection{Agnosticism against unseen signal models}
When building a discriminator with the ADNN approach one needs to mind the fact that the model chosen by Nature is unknown and in general different from the ones the discriminator has been trained on. This is indeed the primary reason why one wants the discriminator performance to be model independent in the first place. It is thus important to establish a procedure to evaluate the degree of model independence of the discriminator with respect to signal models unseen in the training. In the case under study one may expect that an unknown mixture between CP-even and CP-odd couplings is found in Nature. A check of the model independence against a generic mixture that was not used in the training is needed to assess the residual model dependence.

\begin{figure}[ht]
    \centering
    {\includegraphics[width=.455\columnwidth]{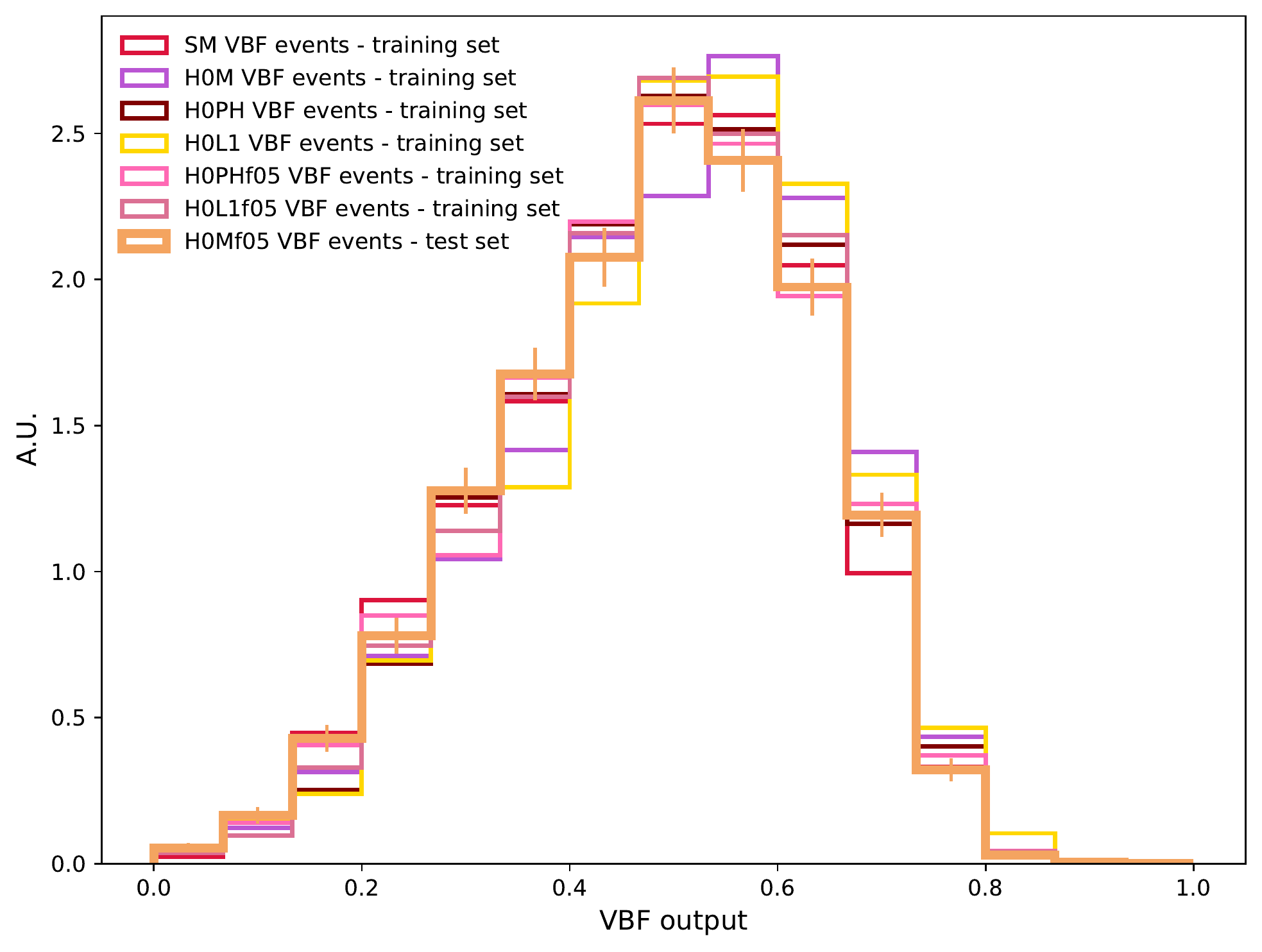}}
    \caption{\emph{Normalized distribution of the VBF-output of an ADNN trained excluding H0Mf05 in C1. The uncertainty associated with the limited size of simulated samples is reported only in the H0Mf05 distribution.}}
    \label{ADNN_excl1}
    \end{figure}

As an application of this procedure in the \hww case study, an ADNN was trained excluding one of the mixed models and compared the discriminator shape between the models used in the training and the excluded one. Results obtained in C1 are shown in Figure~\ref{ADNN_excl1}, where the discriminator shape for the model excluded from the training (ticker line) is compared to that of those models that were used instead. In this case, having trained against the pure models is sufficient for the discriminator to be agnostic against the mixed model.

This result may be peculiar to the case under study, but it is important to stress that it is always possible to test the ADNN performance against unseen models, and, in case of unsatisfactory results, enrich the set of models in the training to achieve the desired level of model independence. This is particularly interesting when paired with the fact that any deviation from the SM can be modeled in the effective field theory approach by a finite set of operators, and that one can, at least in principle, train a discriminator to be agnostic with respect to all of them (or, more realistically, all the ones that have an influence on the observables that are being measured). 

\subsection{Signal extraction and bias estimation}\label{sec:bias_estimation}
The attention is now turn to the estimation of the residual bias in the measurement, to quantify the reduction made possible by the ADNN approach presented in this paper.

The signal extraction is performed through a binned maximum likelihood fit. The number of observed data $N_{\mathrm{i,obs}}$ in the $i$-th bin is expected to follow a Poisson distribution with mean value ($\mu s_\mathrm{i} + b_\mathrm{i}$), where $s_\mathrm{i}$ and $b_\mathrm{i}$ are the expected signal and background yields in the bin, respectively. The signal strength modifier $\mu$ is introduced to check the agreement between the measured number of signal events and the prediction of the considered physics hypothesis, and is defined as the ratio between the measured signal cross section and the expected one: \begin{equation}
    \sigma = \mu \  \sigma^{\mathrm{exp}}.
\end{equation}
To take any systematic error affecting the MC simulation into account, each source of uncertainty is modeled as an individual nuisance parameter $\nu$. The likelihood function is then expanded by including a corresponding constraint $\mathcal{N}(\boldsymbol{\nu})$. Only the uncertainties related to the signal and background processes theoretical modeling i.e., scale and parton distribution function uncertainties, have been considered.\\
In conclusion, the likelihood function used for the template fit is: \begin{equation}\label{L-1}
    L(\mu,\boldsymbol{\nu}) =  \prod_{\mathrm{i}=1}^N \mathcal{P} \left( N_{\mathrm{i,obs}}\ ; \mu s_\mathrm{i}(\boldsymbol{\nu}) + b_\mathrm{i}(\boldsymbol{\nu})\right)\cdot\mathcal{N}(\boldsymbol{\nu}).
\end{equation}
Although the nuisance parameters are determined through the fit procedure, they are not relevant for the analysis purpose and the only parameter of interest is the signal strength modifier $\mu$.

With the goal of comparing the performance of the ADNN and the DNN trained on the SM only, the uncertainties on the measured VBF cross section was evaluated by employing the VBF-output of the two types of networks.
To do so, only the SM contribution was considered as the signal process. A data set was constructed by generating the observed numbers of events in each bin such that they are equal to the sum of the expected signal and background events and with a Poisson uncertainty. This leads to a pseudo data set matching exactly the sum of all MC histograms, in which the value of the parameter of interest $\mu$ returned by the template fit is equal to $1$ by construction. A data set with these properties is called an Asimov data set. Therefore, as a first step, the fit procedure is carried out on the Asimov data set corresponding to the assumption of SM cross sections and by using the template distribution of the VBF-output of the classifier both in C1 and C2. The Table~\ref{results_mu} summarizes the uncertainties in the signal strength parameters, corresponding to $68\%$ Confidence Level (CL) intervals, obtained considering the VBF production mechanism as the signal process. The uncertainty associated with the limited size of MC samples is roughly 23-27$\%$ (25$\%$) of the total uncertainty in C1 (C2) using both the ADNN and DNN, whereas the systematic contribution is  the 20-40$\%$ (10-15$\%$) of the total uncertainty in C1 (C2). The total signal strength uncertainty observed in the two categories when the ADNN is used is found to be comparable with the DNN one within the MC sample size uncertainties. Therefore, the adversarial term does not lead to a sizable worsening of the performance. This is expected since the ADNN and the DNN have similar power in discriminating signal from backgrounds.\\
\begin{table}[ht]
    \centering
    \caption{\emph{Expected $68\%$ CL uncertainties (statistical and systematic) in the signal strength parameters, evaluated on the Asimov data set corresponding to the SM hypothesis and employing the VBF-output of both the ADNN and the DNN as a fit variable.}}
    \begin{tabular}{ccc}
    \toprule
    & DNN & ADNN \\ \midrule
    Category &  Total uncertainty &  Total uncertainty\\ \midrule
    C1 & $-0.38 / +0.55 $ &  $-0.39 / +0.56 $\\
    C2 & $-0.17 / +0.20 $ &  $-0.20 / +0.22 $\\
    \bottomrule
    \end{tabular}
\label{results_mu}
\end{table}

\begin{figure*}[ht]
    \centering
    {\includegraphics[width=0.8\columnwidth]{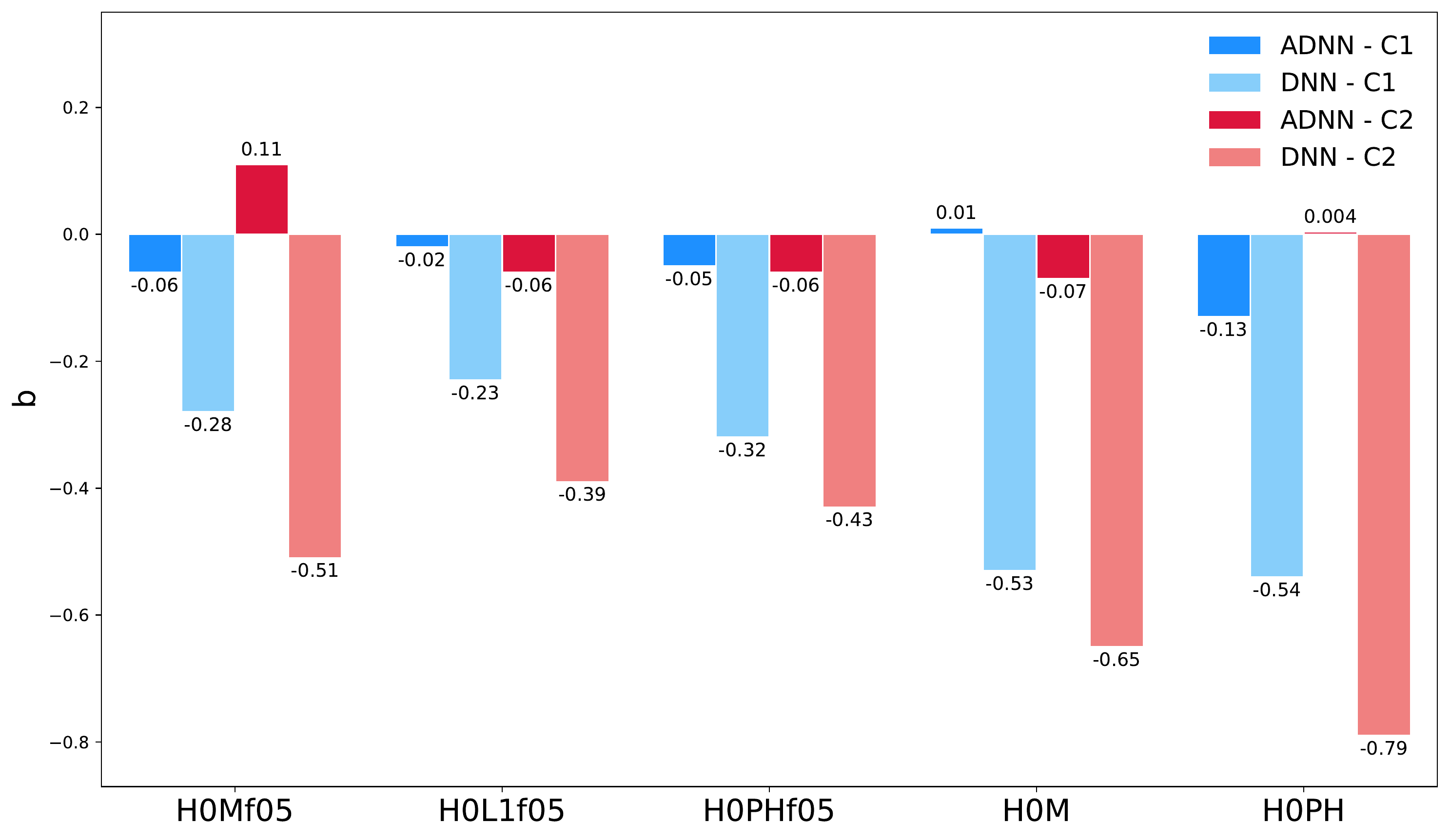}} 
    \caption{\emph{Resulting bias quantities evaluated with both the DNN and the ADNN in the C1 and C2 categories.}}
    \label{histo_bias}
    \end{figure*}

\begin{figure}[ht]
    \centering
    \subfloat[C1]
    {\includegraphics[width=.45\columnwidth]{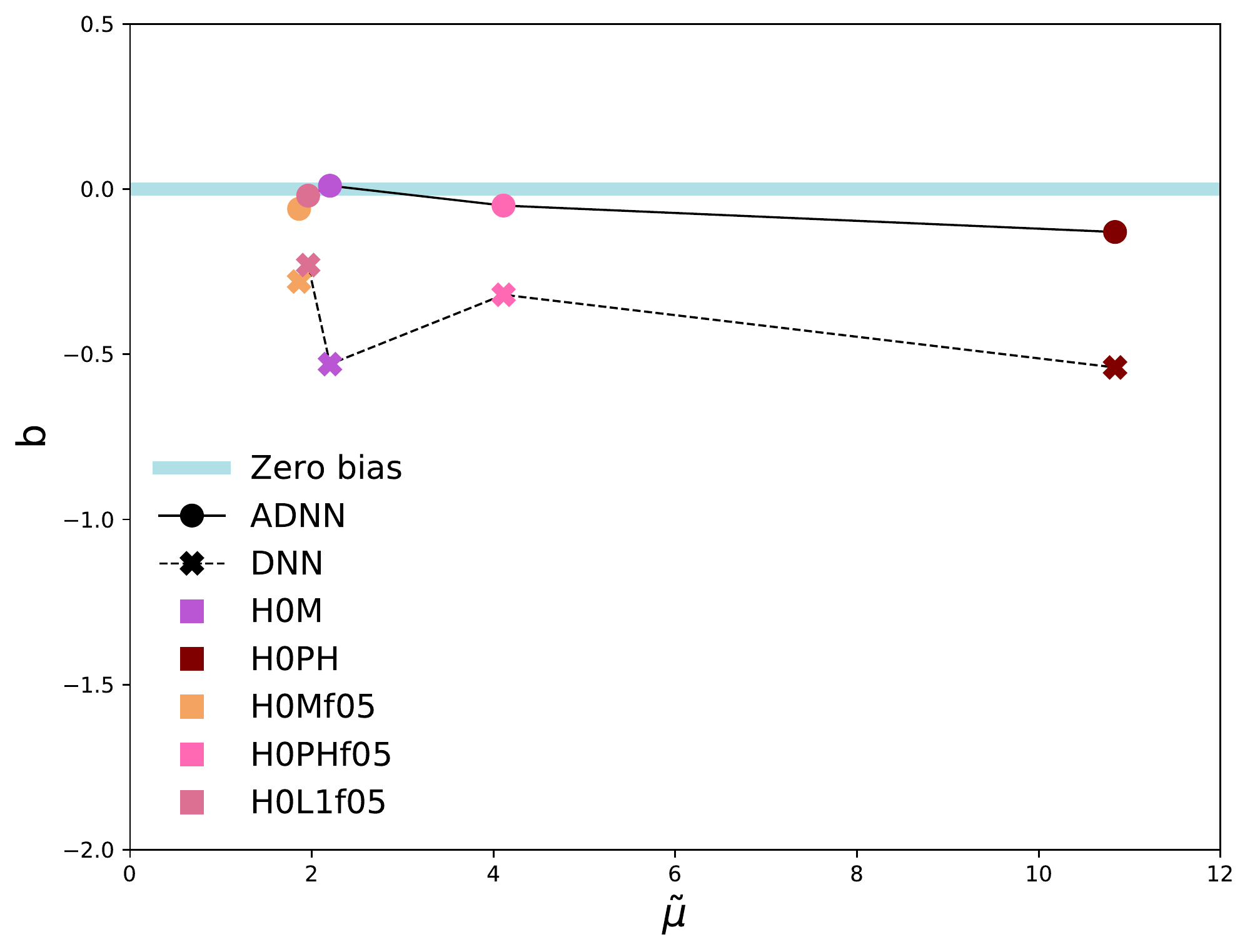}}
    \subfloat[C2]
    {\includegraphics[width=.45\columnwidth]{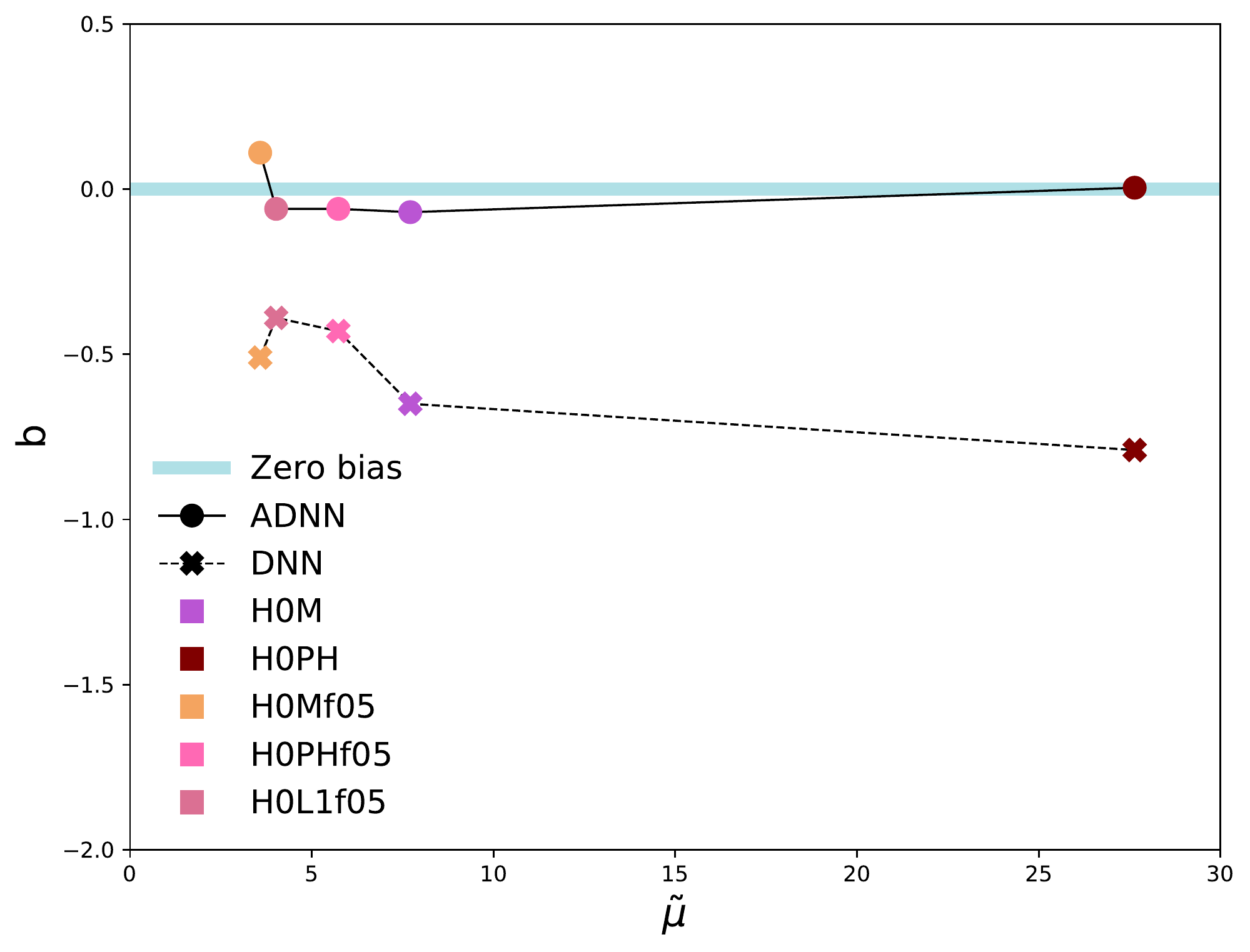}} \\
    \caption{\emph{Bias quantities as a function of the expected value of the signal strength modifier.}}
    \label{bias_c1_c2}
    \end{figure}

Subsequently, an Asimov pseudo-data set is generated assuming one of the considered signal BSM hypothesis and all background contributions. This data set is then fitted using the SM signal and background MC templates.
For each category, the fit provides the estimator $\hat{\mu}$ which is equal to the number of measured BSM signal events over the number of expected SM signal events:
\begin{equation}\label{muj}
    \hat{\mu} = \frac{s_{\mathrm{BSM}}^{\mathrm{fit}}}{s_{\mathrm{SM}}^{\mathrm{exp}}}
\end{equation}
In this particular case of study, $\hat{\mu}$ corresponds to the ratio between the measured BSM cross section and the expected SM one. Let $\tilde{\mu}$ be defined as the ratio of the BSM and SM cross sections:  \begin{equation}
    \tilde{\mu} = \frac{\sigma_{\mathrm{BSM}}}{\sigma_{\mathrm{SM}}}
\end{equation}
Therefore, $\tilde{\mu}$ is the signal strength that one would expect to measure if there was no bias due to the shape effect in the fit procedure. The (B)SM cross section is given by the number (B)SM events that enter in the considered category $N_{\mathrm{(B)SM}}$ divided by the integrated luminosity $\mathcal{L}$.
Finally, the total bias as a fraction of the expected value $\tilde{\mu}$ is introduced: \begin{equation}
    b =\frac{\hat{\mu} - \tilde{\mu}}{\tilde{\mu}} 
\end{equation}
The bias quantities estimated with the DNN trained on SM events and the ADNN are shown in Table~\ref{table_SMeCo_bias} for all the BSM hypotheses except the H0L1. Given its small cross section, no events generated according to the H0L1 model enter the analysis phase space.\\ As already said, these BSM models describe HVV couplings that are extremely different from the SM case and this is also outlined by the fact that the fit provides values of the $\hat{\mu}$ parameter that are much higher than one.\\
Using the ADNN for the analysis procedure allows to strongly reduce the bias quantities in both the C1 and C2 regions with respect to the case where the DNN is employed. This behavior is expected since, as illustrated in Figure~\ref{DNN_VBF_output}, distributions of the VBF-output of the DNN on all the BSM signal events differ greatly from the one on SM signal events. In quantitative terms, the DNN bias is roughly 20-80$\%$ of the expected value $\tilde{\mu}$, depending on the BSM theory used to generate the Asimov data set and on the category in which the analysis is carried out. The C2 category shows model dependence values larger than the ones affecting the C1 region, but this is justified by the fact that the difference in the VBF-output shape between SM and BSM events is even more pronounced when compared to the C1 category. On the other hand, the ADNN makes it possible to reduce the bias to less than 10$\%$ of $\tilde{\mu}$ for almost all the BSM models. Moreover, the ADNN biases are comparable in magnitude with respect to the statistical uncertainty on $\hat{\mu}$ related to the finite size of the MC samples, whereas for the DNN case biases are 5 to 10 times larger than such uncertainties.  For what the systematic component is concerned, in both the categories it spans from 75$\%$ to 95$\%$ of the total error depending on the BSM model used to generate the Asimov template.
\\
Same results are also summarized in Figure~\ref{histo_bias}, where the bias estimated using each BSM theories is reported for both the C1 and C2 categories. 
\begin{table}[ht]
\centering   
    \caption{\emph{Expected and measured signal strength and bias quantities evaluated on the Asimov data set corresponding to each BSM hypothesis with both the DNN and the ADNN.}}
    \begin{tabular}{cc|cc|cc}
    \toprule
    
    \multicolumn{6}{c}{\textbf{H0Mf05}}\\\toprule
    & &  \multicolumn{2}{c}{DNN} & \multicolumn{2}{|c}{ADNN}\\\midrule
    $Category$& $\tilde{\mu}$ &  $\hat{\mu}$ & $b$ & $\hat{\mu}$ & $b$ \\ \midrule
    C1 & $1.86$ &  $1.34$ & $-0.28$ & $1.75$ & $-0.06$\\
    C2 & $3.58$ &  $2.77$ & $-0.51$ & $3.17$ & $0.11$ \\
    \hline
    \multicolumn{6}{c}{} \\\toprule
    
    \multicolumn{6}{c}{\textbf{H0L1f05}}\\\toprule
    & &  \multicolumn{2}{c}{DNN} & \multicolumn{2}{|c}{ADNN}\\\midrule
    $Category$& $\tilde{\mu}$ &  $\hat{\mu}$ & $b$ & $\hat{\mu}$ & $b$ \\ \midrule
    C1 & $1.96$ &  $1.52$ & $-0.23$ & $1.92$ & $-0.02$\\
    C2 & $4.02$ &  $2.46$ & $-0.39$ & $3.76$ & $-0.06$ \\
    \hline
    \multicolumn{6}{c}{} \\\toprule
    
    \multicolumn{6}{c}{\textbf{H0PHf05}}\\\toprule
    & &  \multicolumn{2}{c}{DNN} & \multicolumn{2}{|c}{ADNN}\\\midrule
    $Category$& $\tilde{\mu}$ &  $\hat{\mu}$ & $b$ & $\hat{\mu}$ & $b$ \\ \midrule
    C1 & $4.11$ &  $1.34$ & $-0.32$ & $3.91$ & $-0.05$\\
    C2 & $5.73$ &  $1.77$ & $-0.43$ & $5.39$ & $-0.06$ \\
    \hline
    \multicolumn{6}{c}{} \\\toprule
    
    \multicolumn{6}{c}{\textbf{H0M}}\\\toprule
    & &  \multicolumn{2}{c}{DNN} & \multicolumn{2}{|c}{ADNN}\\\midrule
    $Category$& $\tilde{\mu}$ &  $\hat{\mu}$ & $b$ & $\hat{\mu}$ & $b$ \\ \midrule
    C1 & $2.20$ &  $1.04$ & $-0.53$ & $2.22$ & $0.01$\\
    C2 & $7.71$ &  $2.69$ & $-0.65$ & $7.13$ & $-0.07$ \\
    \hline
    \multicolumn{6}{c}{} \\\toprule
    
    \multicolumn{6}{c}{\textbf{H0PH}}\\\toprule
    & &  \multicolumn{2}{c}{DNN} & \multicolumn{2}{|c}{ADNN}\\\midrule
    $Category$& $\tilde{\mu}$ &  $\hat{\mu}$ & $b$ & $\hat{\mu}$ & $b$ \\ \midrule
    C1 & $10.84$ &  $4.98$ & $-0.54$ & $9.40$ & $-0.13$\\
    C2 & $27.64$ &  $5.82$ & $-0.79$ & $27.76$ & $0.004$ \\
    \hline
    \multicolumn{6}{c}{} \\\toprule
    \end{tabular}
    \label{table_SMeCo_bias}
    \end{table}
\\Figure~\ref{bias_c1_c2} shows the resulting bias quantities as a function of the expected signal strength modifier value, estimated with both the DNN and the ADNN in each of the two categories. It can be noticed that the bias is a roughly constant fraction of the $\tilde{\mu}$ parameter when the ADNN or the DNN is employed, and this fraction is decreased to values below 10$\%$ through the usage of the ADNN in both the categories.

\section{Conclusion}
In this work, an implementation of the domain adaptation technique for defining a learning algorithm that is independent on the data sample on which it is trained has been proposed. The implementation is based on a system of two neural networks trained in a competitive way with an adversarial technique (ADNN).\\
To confirm the effectiveness of this approach, it was applied to the use case of the \hww STXS cross section measurement at the LHC with the aim to reduce the model dependence of the final results introduced by the signal extraction procedure. The measurement targets the VBF process as the signal process and it has been performed at particle-level, employing a simulated data sample from p-p collisions corresponding to an integrated luminosity of 138 fb$^{-1}$. Compared to a standard feed-forward deep neural network, the usage of the ADNN allowed the measurement of the VBF cross section with the same level of precision, but significantly reducing the measurement bias due to the signal modeling assumptions.

\section*{Acknowledgements} 
The authors acknowledge the CMS Collaboration, in particular the computing, Monte Carlo, and $\mathrm{H\to WW}$ working groups, for providing the computing resources used for the Monte Carlo event simulation of the physics processes studied in this paper. 

\bibliography{main} 
\bibliographystyle{ieeetr}

\end{document}